\documentclass[a4paper,fleqn,usenatbib]{mnras}

\usepackage[T1]{fontenc}
\usepackage{ae,aecompl}

\usepackage{graphicx}          
\usepackage{amsmath}           
\usepackage{amssymb}           
\usepackage{times}             
\usepackage{enumitem}          
\usepackage{chngcntr}
\usepackage{xcolor}

\pdfoutput=1

\newcommand{\MJ}{\mbox{$M_{J}$}\,}

\newcommand{\tx}{\mbox{$\times10$}}

\newcommand{\UG}{\mbox{$u-g$}\,}        
\newcommand{\GR}{\mbox{$g-r$}\,}
\newcommand{\GI}{\mbox{$g-i$}\,}
\newcommand{\RI}{\mbox{$r-i$}\,}
\newcommand{\RZ}{\mbox{$r-z$}\,}
\newcommand{\IZ}{\mbox{$i-z$}\,}
\newcommand{\ZJ}{\mbox{$z-J$}\,}
\newcommand{\VJ}{\mbox{$V-J$}\,}
\newcommand{\JH}{\mbox{$J-H$}\,}
\newcommand{\JK}{\mbox{$J-K_S$}\,}
\newcommand{\HK}{\mbox{$H-K_S$}\,}
\newcommand{\JWa}{\mbox{$J-W1$}\,}
\newcommand{\JWb}{\mbox{$J-W2$}\,}
\newcommand{\HWa}{\mbox{$H-W1$}\,}
\newcommand{\HWb}{\mbox{$H-W2$}\,}
\newcommand{\KWb}{\mbox{$K_S-W2$}\,}

\title[Selecting M dwarfs with Unresolved UCD Companions]{A Method for Selecting M dwarfs with an Increased Likelihood of Unresolved Ultra-cool Companionship}
\author[N. J. Cook et~al.]{N. J. ~Cook,\thanks{E-mail: \href{mailto:neil.james.cook@gmail.com}{neil.james.cook@gmail.com}}$^{1}$
D. J. ~Pinfield,$^{1}$ F. ~Marocco,$^{1}$ B. ~Burningham,$^{1, 2}$ H. R. A. ~Jones,$^{1}$ 
\newauthor
J. ~Frith,$^{1}$ J. ~Zhong,$^{3}$ A. L. ~Luo,$^{4}$ Z. X. ~Qi,$^{3}$ P. W. ~Lucas,$^{1}$ M. Gromadzki,$^{5, 6}$ 
\newauthor
A. C. ~Day-Jones,$^{1}$ R. G. ~Kurtev,$^{6, 5}$ Y. X. ~Guo,$^{4}$ Y. F. ~Wang,$^{4}$ Y. ~Bai,$^{4}$ Z. P. ~Yi,$^{7}$
\newauthor
and R. L. ~Smart$^{8}$
\\\\
$^{1}$Centre for Astrophysics Research, Science and Technology Research Institute, University of Hertfordshire, Hatfield AL10 9AB, UK \\
$^{2}$NASA Ames Research Center, Mail Stop 245-3, Moffett Field, CA 94035, USA\\
$^{3}$Key Laboratory for Research in Galaxies and Cosmology, SHAO, Chinese Academy of Sciences, 80 Nandan Road, Shanghai 200030, China\\
$^{4}$Key Laboratory of Optical Astronomy, NAO, Chinese Academy of Sciences, Datun Road 20A, Beijing 100012, China\\
$^{5}$Millennium Institute of Astrophysics, Av. Vicua Mackenna 4860, 782-0436, Macul, Santiago, Chile\\
$^{6}$Istitiuto de F{\'i}sica y Astronom{\'i}a, Universidad de Valpara{\'i}so, ave. Gran Breta\~{n}a, 1111, Casilla 5030, Valpara{\'i}so, Chile\\
$^{7}$Shandong University at Weihai,  Weihai, 264209, China\\
$^{8}$Istituto Nazionale di Astrofisica - Osservatorio Astrofisico di Torino, Via Osservatorio 20, I-10023, Torino\\
}

\date{Accepted 2016 January 6. Received 2015 August 17}

\pubyear{2016}

\begin{document}
\label{firstpage}
\pagerange{\pageref{firstpage}--\pageref{lastpage}}

\include{aas_macros}

\maketitle

\begin{abstract}
    Locating ultra-cool companions to M dwarfs is important for constraining low-mass formation models, the measurement of sub-stellar dynamical masses and radii, and for testing ultra-cool evolutionary models. We present an optimised method for identifying M dwarfs which may have unresolved ultra-cool companions. We construct a catalogue of 440,694 M dwarf candidates, from WISE, 2MASS and SDSS, based on optical and near-infrared colours and reduced proper motion. With strict reddening, photometric and quality constraints we isolate a sub-sample of 36,898 M dwarfs and search for possible mid-infrared M dwarf + ultra-cool dwarf candidates by comparing M dwarfs which have similar optical/near-infrared colours (chosen for their sensitivity to effective temperature and metallicity). We present 1,082 M dwarf + ultra-cool dwarf candidates for follow-up. Using simulated ultra-cool dwarf companions to M dwarfs, we estimate that the occurrence of unresolved ultra-cool companions amongst our M dwarf + ultra-cool dwarf candidates should be at least four times the average for our full M dwarf catalogue. We discuss possible contamination and bias and predict yields of candidates based on our simulations.
\end{abstract}

\section{Introduction}
    \label{section:intro}

    The ultra-cool ($T_{eff}$<2500 K, >M7) field population has been greatly expanded over the last 15 years using large-scale red and infrared surveys; {\itshape The Two Micron All-Sky Survey} \citep[2MASS,][]{Skrutskie2006}, {\itshape The Sloan Digital Sky Survey} \citep[SDSS,][]{York2000}, {\itshape The UKIRT Infrared Deep Sky Survey} \citep[UKIDSS,][]{Lawrence2007}, {\itshape The Visible and Infrared Survey Telescope for Astronomy} \citep[VISTA,][]{Emerson2002} and {\itshape The Wide-Field Infrared Survey Explorer} \citep[WISE,][]{Wright2010}. Evolutionary cooling means the masses of these sources are age-sensitive, ranging from old low-mass stars through younger brown dwarfs (e.g. in \citealt{Nakajima1995}, \citealt{Delfosse1997}, \citealt{Burgasser1999}, \citealt{Kirkpatrick1999}, \citealt{Pinfield2003}, \citealt{Burgasser2004}, \citealt{Leggett2010}, and \citealt{Kirkpatrick2011}) and down into the planetary regime (e.g. in \citealt{Lucas2006}, \citealt{Caballero2007}, \citealt{Luhman2008}, \citealt{Marsh2010}, \citealt{Lodieu2011}, \citealt{Delorme2012}, and \citealt{Scholz2012}).

    Ultra-cool dwarfs (UCDs) can be extremely informative. Statistical studies of UCD companions aid the theoretical study of low-mass star formation and provide constraints on the initial mass function \mbox{\citep{Parker2013,Chabrier2014}}. Specifically, companion statistics can be used to decide between different formation processes. For wide binaries, e.g., it is difficult to explain formation via dynamical processes or disc fragmentation with such systems possibly forming via early stage core fragmentation \citep{Chabrier2014}. There is also a lack of 10 to 100 Jupiter mass companions (the brown dwarf desert) in separation ranges covered by radial velocity surveys. Observationally, the frequency of companions increases for planetary-mass objects but decreases for objects with larger mass \citep{Howard2010}. Giant planets are also less frequent around lower-mass stars than higher mass stars \citep{Johnson2010} whereas, in the same separation range, brown dwarf companions become more frequent around low-mass stars and other brown dwarfs \citep{Joergens2008}. 

    Ultra-cool companions are also useful as benchmark sources to test structure and atmospheric evolutionary models \citep{Pinfield2006}. Companions where physical parameters can be directly measured can be used for testing atmospheres and structure models of both the primaries and companions \citep{Baraffe2003,Burrows2011,Luhamn2012,Allard2012,Saumon2012}. Companion properties such as age and composition can generally be inferred from the primary star \citep{Leggett2010}, and mass and radius constraints can come from radial velocity and light curve studies over multiple orbital periods \citep[e.g.][]{Agol2005,Cumming2008,Jones2015}, via astrometry \citep[e.g. with Gaia,][]{Bruijne2012}, or via adaptive optics \cite[e.g. ][]{Dupuy2010}.

    Unresolved companions have been identified using a variety of observational techniques. High contrast systems are generally revealed through radial velocity variability, with much lower contrast systems (e.g. late M dwarf+UCD or UCD+UCD systems \citealt{Burgasser2006b}, \citeyear{Burgasser2010}; and \citealt{Gagliuffi2013}, \citeyear{BardalezGagliuffi2015}) more amenable to spectroscopic and photometric study (e.g. in \citealt{Reid2000}, \citealt{Reid2001}, \citealt{Oppenheimer2001}, \citealt{Nidever2002}, \citealt{Pinfield2003}, \citealt{Burgasser2003}, \citealt{Close2003}, \citealt{Reiners2004}, \citealt{Burgasser2006}, \citealt{Joergens2008}, \citealt{Luhamn2012}, and \citealt{Todorov2014}).

    Low number statistics and observational bias make it difficult to robustly constrain the M dwarf companion fraction with varying mass-ratio and separation. It is clear however, that for separations of $\la$100 AU the M dwarf+UCD companion fraction is at the level of approximately one per cent (i.e. 2-4 per cent, via Adaptive optics; \citealt{Neuhauser2004}; 0-2 per cent, 0.001<$\theta$<0.01 AU, \citealt{Reid2000}; 1-3 per cent, 10<$\theta$<100 AU \citealt{Oppenheimer2001}; 1 per cent, 0.1<$\theta$<1 AU, \citealt{Nidever2002}; and 1 per cent, 1.0<$\theta$<10.0 AU \citealt{Nidever2002}, where $\theta$ is separation).

    In this paper we present a new photometric method which aims to provide significant benefits to targeted searches for unresolved UCD companions to M dwarfs. We take advantage of the extensive multi-band photometry (optical to mid-infrared) available by combining the WISE, 2MASS and SDSS surveys, and the large M dwarf sample size this reveals (Section \ref{section:Catalogue_selection}). By constructing a large catalogue of well measured M dwarfs in un-reddened regions of the sky, we isolate a sample which can be searched for mid-infrared outliers within a multi-colour parameter-space. We then optimise a method to search for such outliers, which could be due to unresolved UCD companions, by carefully minimising the colour variation expected from effective temperature and metallicity differences. We interpret our results via simulations, by artificially injecting an unresolved companion population into our sample, and assessing the increased likelihood of such systems appearing in different regions of the multi-colour parameter-space. This allows us to select a sample of M dwarfs with significantly increased potential for unresolved UCD companionship (Section \ref{section:selecting_Ms_with_MIR_excess}). We present our candidates, discuss possible contamination and bias, and predict yields and the expected companion spectral type distribution based on our simulations. We summarise our results and discuss planned follow-up in Section \ref{section:summary}.

\section{Catalogue selection}
    \label{section:Catalogue_selection}

    As a foundation for our analysis procedures, we construct a catalogue of M dwarf candidates with high quality WISE/2MASS/SDSS photometry. We chose not to use SDSS $u$ band (0.3551 $\mu$m) due to its increased uncertainties \citep{Padmanabhan2008}, and  chose not to use the WISE $W3$ and $W4$ bands due to the greatly reduced sensitivity \citep[Magnitude limit at 5$\sigma$ of 11.40 for the 12 $\mu$m, $W3$ band, and 7.97 and for the 22 $\mu$m, $W4$ band; ][]{Wright2010} as we would not have detections for many of our M dwarf candidates.

    \begin{table*}
        \begin{center}
        \begin{tabular}{p{1cm}p{1.5cm}p{1.5cm}p{1.5cm}p{2.5cm}p{0.5cm}}
        \hline
        Survey & Band & Wavelength & PSF-FWHM & Magnitude-limit (5$\sigma$) & Notes \\ 
         & & $\mu$m & $arcsec$ & $mag$ & \\
        \hline
        SDSS & $g$ & 0.4686 & 1.3 & 22.2 & $a$ \\
        SDSS & $r$ & 0.6165 & 1.3 & 22.2 & $a$ \\
        SDSS & $i$ & 0.7481 & 1.3 & 21.3 & $a$ \\
        SDSS & $z$ & 0.8931 & 1.3 & 20.5 & $a$ \\

        2MASS & $J$   & 1.25 & 2.9 & 16.55 & $b$ \\ 
        2MASS & $H$   & 1.65 & 2.8 & 15.85 & $b$ \\ 
        2MASS & $K_S$ & 2.16 & 2.9 & 15.05 & $b$ \\ 

        WISE & $W1$ & 3.4 & 6.1 & 16.5 & $c$ \\
        WISE & $W2$ & 4.6 & 6.4 & 15.5 & $c$ \\
        \hline
        \multicolumn{6}{|p{0.7\textwidth}|}{$^a$ \protect\citet{Ahn2012} and \protect{\url{http://www.sdss3.org/dr9/scope.php}}} \\
        \multicolumn{6}{|p{0.7\textwidth}|}{$^b$ \protect\citet{Skrutskie2006}, \protect{\url{http://www.ipac.caltech.edu/2mass/releases/allsky/} and \protect{\url{http://spider.ipac.caltech.edu/staff/roc/2mass/seeing/seesum.html}}} (Magnitude limits quoted as 10$\sigma$, 5$\sigma \equiv$ `mag at 10$\sigma$' + 0.75)}\\
        \multicolumn{6}{|p{0.7\textwidth}|}{$^c$ \protect\citet{Wright2010} and \protect{\url{http://wise2.ipac.caltech.edu/docs/release/allsky/expsup/}}}\\
    	\end{tabular}
    	\end{center}
        \caption{Summary of SDSS \protect\citep{Ahn2012}, 2MASS protect\citep{Skrutskie2006} and WISE \protect\citep{Wright2010} photometric bands used in our selection process. \label{table:phot_summary}}      
    \end{table*}

    \subsection{Initial colour and photometric cuts}
    	\label{section:Catalogue_selection:colour_cuts}
       
        We began by downloading\footnote{Access to data releases via \url{http://irsa.ipac.caltech.edu} \label{footnote:IRSA}} all 563,921,603 sources in the WISE All-Sky catalogue, of which 280,909,458 had 2MASS counterparts within three arcsec. Table \ref{table:phot_summary} summarises the multi-band photometric sensitivity limits that we used when selecting sources from these surveys. We applied near-infrared colour cuts to help remove contaminating giants and earlier spectral type stars from our sample. We made use of the \JH and \HK colour constraints from \citet[][hereafter LG11]{Lepine2011}. After these initial cuts 57,510,435 sources remained.

        We then cross-matched the catalogue with {\itshape The Tenth Data Release of SDSS}\footnote{Access via \url{http://skyserver.sdss3.org/CasJobs/} \label{footnote:SDSS}}\citep[SDSS DR10]{Ahn2012}. Of the 57 million sources, 9,944,123 sources had SDSS photometry and were flagged as stars ($type=6$ in DR10 {\itshape PhotoObjAll}). Using the \GR to $V-g$ transformation\footnote{SDSS photometric transformations available via \url{https://www.sdss3.org/dr8/algorithms/sdssUBVRITransform.php}} of \citet{Jordi2006}, we were able to calculate an estimate of the V band magnitudes. This enabled a spectral type estimation via \VJ \citep[see equation 12; ][]{Lepine2013}. We removed all sources which have an estimated spectral type early than $\sim$M3.5 (equivalent to a \VJ>4.0, see Section \ref{section:mir_excess:excess_signatures}). This is a slightly redder cut than (LG11, which aimed to remove stars earlier than K7 dwarfs with \VJ>2.7). This left 1,352,931 sources with estimated spectral type M3.5 or later. 

        To reduce the number of sources which had poor photometry we used basic accuracy cuts of 0.1 in the uncertainties of $V$, $J$, $H$, $K$, $W1$ and $W2$. This left a total of 704,723 sources. We impose more strict photometric requirements for the excess analysis in Section \ref{section:Catalogue_selection:reddeneing}.

    \subsection{Reduced proper motion}
    	\label{section:Catalogue_selection:reducedpm}

        Reliable, accurate proper motions ($\mu$) allows the separation of dwarf stars from background stars and galaxies through reduced proper motion. We therefore cross-matched our sample with the {\itshape Position and Proper Motion Extended-L}$^{\ref{footnote:IRSA}}$ catalogue \citep[PPMXL]{Roeser2010}. A total of 691,421 of the 704,723 sources had proper motion measurements thus we decided not to use any additional proper motion catalogues. Any sources without proper motions in PPMXL were rejected as possible contaminants. We selected only sources whose proper motion uncertainties were <25 per cent (4$\sigma_{\mu}$) of the measured value. Of those sources with proper motion, 464,655 met the 4$\sigma_{\mu}$ cut. Reduced proper motion, $H_V = V + 5log(\mu)+ 5$, was then calculated and possible contamination rejected following the same approach as LG11 ($H_V > 2.2(V-J) + 2.0$). This offers a good balance between contamination rejection and M dwarf retention. LG11 estimated associated M dwarf rejection rates of no more than 1.1 per cent. After the reduced proper motion cut we were left with 450,440 M dwarf candidates. We used the 2MASS proximity flag ({\itshape prox}), to make sure none of our M dwarfs had another 2MASS counterpart within six arcsec. This avoids source blending which can affect photometric accuracy. This left a total of 440,694 sources in our full catalogue of M dwarf candidates.

	%-------------------------------------------------------------------------- 

	\subsection{Catalogue properties}
	    \label{section:the_catalogue}

        \begin{figure*}
            \begin{center}
            \includegraphics[width=\textwidth]{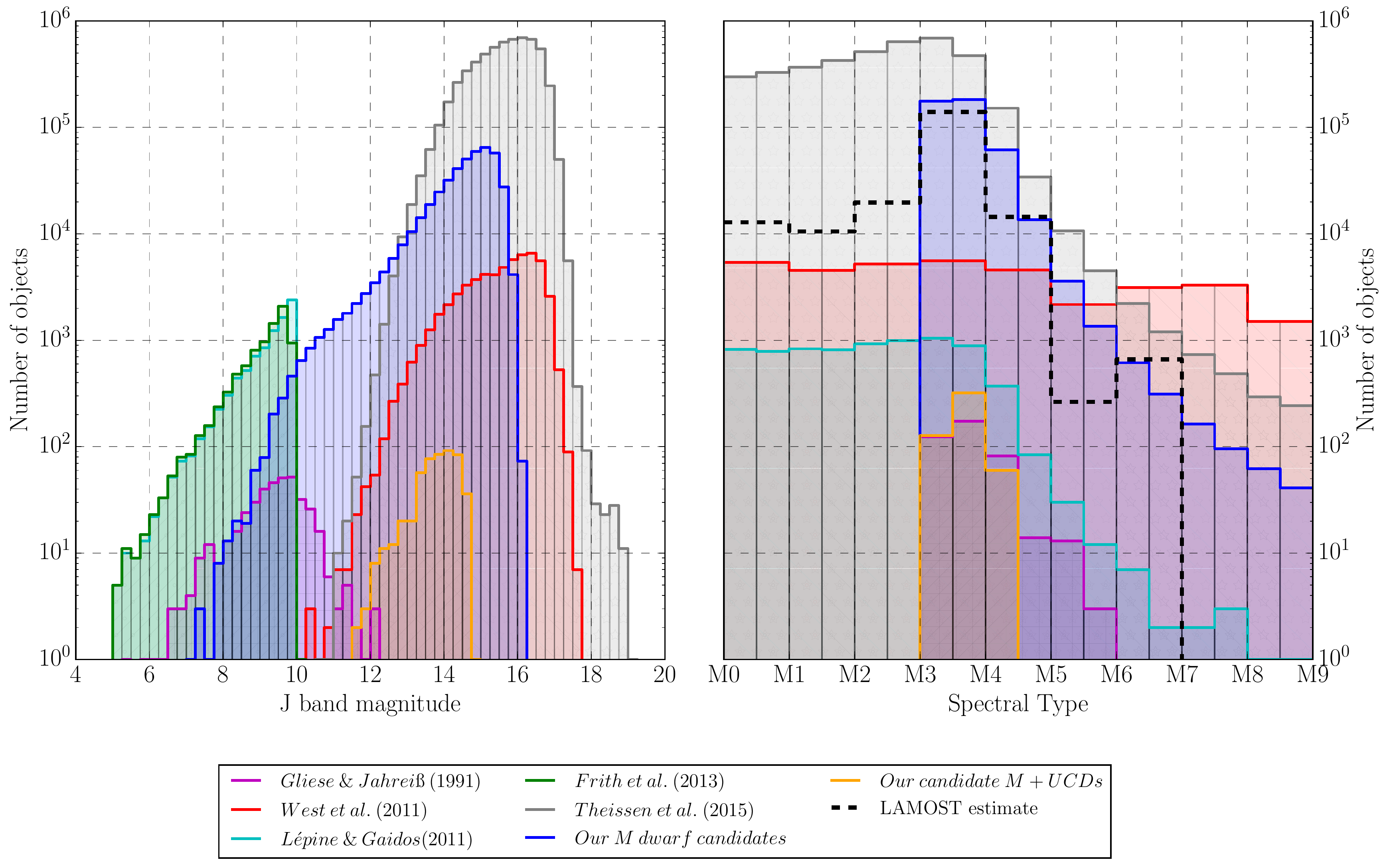}

            \caption{
            Histograms comparing the J band magnitudes (left) and spectral type (right) of our M dwarf candidates and the catalogues of \protect\citet{Gliese1991, West2011}, LG11, \protect\citet{Frith2013} and \protect\citet{Theissen2015}. The dashed black line on the spectral type histogram shows the M dwarf estimates from LAMOST (see section \protect\ref{section:mdwarf_catalogue:the_catalogue:contamination}). Spectral type for our candidates is calculated photometrically (\VJ) by the equations presented in LG11. \protect\citet{Frith2013} does not give a spectral type estimate nor V band magnitudes in order for us to do a \VJ estimation thus we do now have spectral type data for the \protect\citet{Frith2013} catalogue. \protect\citet{West2011} give integer spectral types thus we assume a flat distribution and split these equally between whole and half integer bins (i.e. N M dwarfs in the M0 - M1 bin becomes N/2 M dwarfs in the M0 - M0.5 bin and N/2 M dwarfs in the M0.5 - M1 bin).
                \label{figure:catalogue_compare_j_spt}}
                \end{center}
        \end{figure*}

        The comparison in J magnitude and the spectral type comparison can be seen in Figure \ref{figure:catalogue_compare_j_spt}. Our M dwarf candidates compliment other catalogues of M dwarfs, including M dwarf catalogues from \citet{Gliese1991, West2011}, LG11, \citet{Frith2013} and \citet{Theissen2015}. Our catalogue is not a continuation of the \citet{Frith2013} nor LG11 catalogue due to our use of the SDSS catalogue (thus restricted to the northern hemisphere).

        Our catalogue is brighter than the recent {\it Motion Verified Red Stars (MoVeRS)} catalogue \citep{Theissen2015} due to their cuts in SDSS of $r > 16$. The MoVeRS catalogue also goes two orders of magnitude deeper than our catalogue due to our quality cuts and our requirement of a W2 detection. It should be noted our M dwarfs consist only of M dwarfs later than M3, and this is not true for the other catalogues compared in Figure \ref{figure:catalogue_compare_j_spt}. Our catalogue of M dwarf candidates represents the largest available, given our requirements, filling in the gap in M dwarf candidates between the bright \citet{Frith2013} and LG11 catalogues and the fainter \citet{West2011} and \citet{Theissen2015} catalogues.

        The dashed black line on the spectral type histogram shows the M dwarf estimates from the {\it Large sky Area Multi-Object Fibre Spectroscopic Telescope, LAMOST} (\citealt{Cui2012,Luo2012,Zhao2012}, see section \protect\ref{section:mdwarf_catalogue:the_catalogue:contamination}), the LAMOST estimates shows the \VJ cut does an imperfect job at selecting later than M3 dwarfs, and that the spectral type distribution goes out to at least M7 (although \VJ scatter may suggest a contingent of later types that we have yet to confirm). We select against earlier M dwarfs, and the \citet{West2011} catalogue continues to dominate numerically for the latest spectral type M dwarfs.

        At the bright extreme the M dwarf frequency of our catalogue falls below those of the \citet{Gliese1991}, LG11 and \citet{Frith2013} catalogues, due mainly to our restriction to SDSS sky. Our catalogue dominates numerically in the magnitude range J=10-15.5, but does not go as deep as the \citet{West2011} spectroscopic catalogue.

\begin{figure*}
        \begin{minipage}{.45\textwidth}
            \begin{center}
            \includegraphics[width=\textwidth]{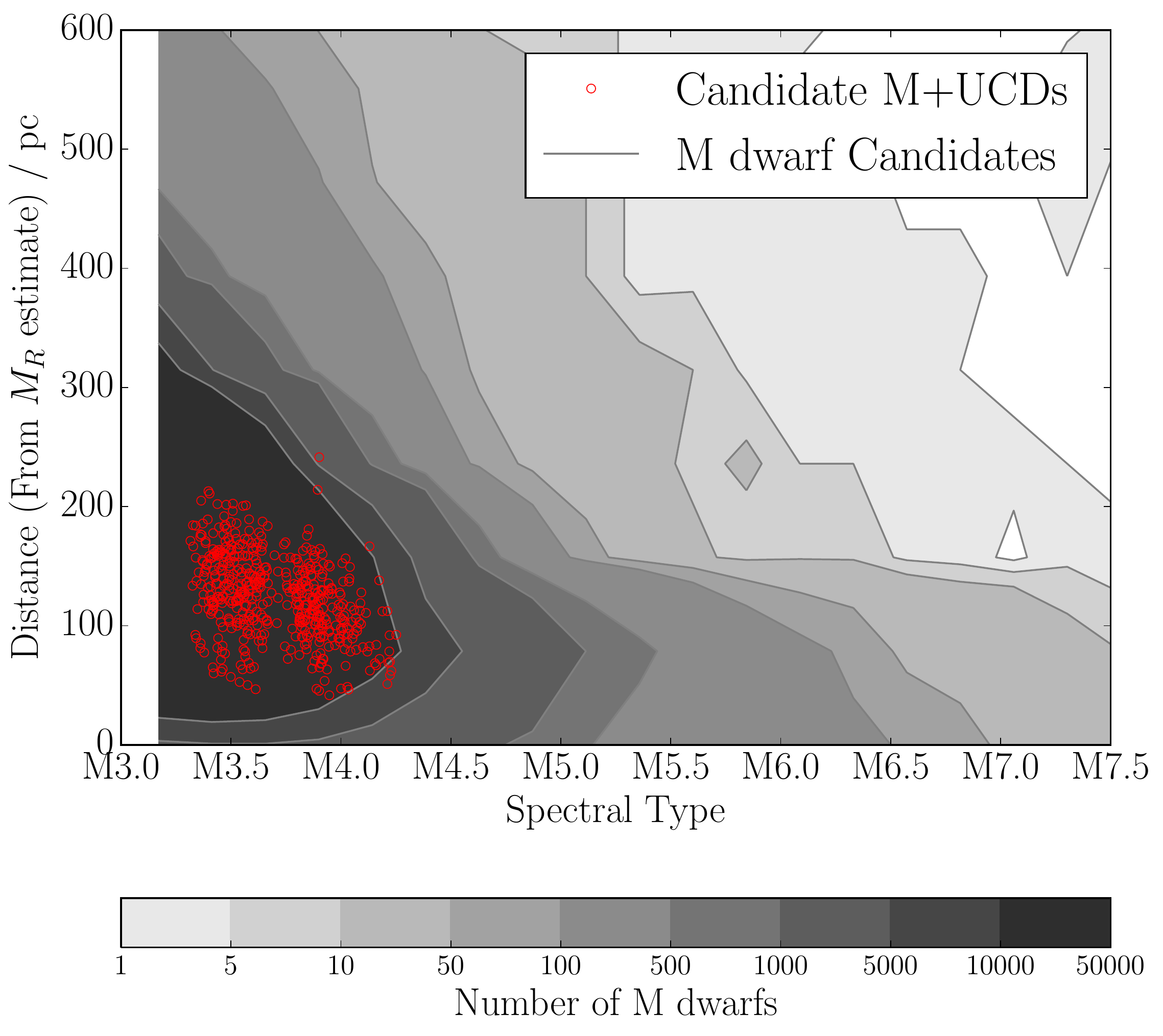}
            \end{center}
            \caption{Distribution in estimated spectral type - distance space. Spectral types from \protect\citet{Lepine2013}'s fit to \VJ. Distance was estimated using \protect\citet{Bochanski2010}'s $M_R$ fits to \RZ and \RI (averaged). \label{figure:spt_vs_distance}}
        \end{minipage}\qquad
        \begin{minipage}{.45\textwidth}
            \begin{center}
                \vspace{0.65cm}
                \includegraphics[width=\textwidth]{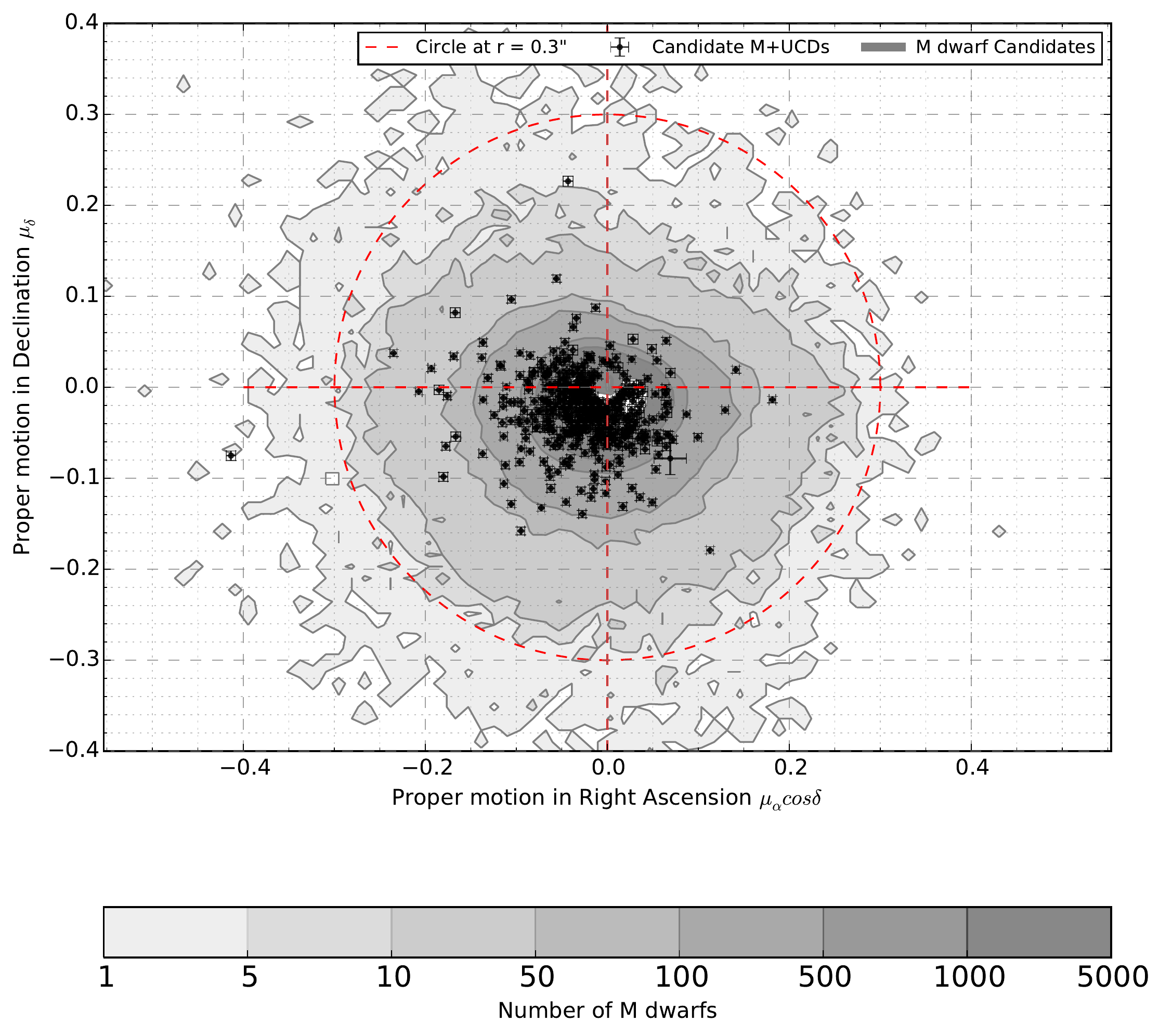}
            \end{center}
            \caption{A proper motion vector-point-diagram showing the overall distribution of M dwarfs in our catalogue as well as our candidate M+UCDs. Distributions are skewed towards the bottom left due to SDSS being a northern hemisphere survey, and in the candidates case being localised around the northern galactic cap. Excess candidates are all within 300 mas yr$^{-1}$ due to the catalogue cross-matching radii selected. \label{figure:pmdiagram}}  
        \end{minipage}
        \end{figure*}

        We estimated distances using the \citet{Bochanski2010} $M_R$ fits to \RZ and \RI. Note these were only used for comparison purposes in Figure \ref{figure:spt_vs_distance}. The bulk of our M dwarf candidates lie between 100 and 200 pc consistent with M dwarfs of spectral type M3 to M5.

    \subsection{Sources of contamination and bias}
        \label{section:mdwarf_catalogue:the_catalogue:contamination}

        We expect our M dwarf candidate catalogue to contain non-M dwarf contamination for two main reasons. Scatter in the \VJ colours will lead to the inclusion of some earlier types (<M3). These will be mostly early M dwarfs but could include some F, G and K stars. Reduced proper motion uncertainty is also expected to lead to a low level of giant contamination as previously discussed.

        To assess the contamination levels we cross-matched our full M dwarf candidate catalogue with the {\itshape Set of Identifications, Measurements and Bibliography for Astronomical Data (SIMBAD)}\footnote{SIMBAD database accessible at \url{http://simbad.u-strasbg.fr/simbad} \citep{Wenger2000} \label{footnote:SIMBAD}} catalogue (cross-matched to three arcsec). In total there were 20,286 matches with our full M dwarf candidate catalogue. Of these 7,360 had spectral types from SIMBAD. From this we gauge our contamination from early (FGK) stars, M giants, and white dwarfs. The full catalogue has $\sim$1.3 per cent contamination from these sources (see Appendix \ref{Appendix:contamination_simbad}). It should however be noted some of the spectral types carry little information, e.g. only as an M-type star ($\sim$1.4 per cent), and thus we may slightly underestimate our contamination from sources such as M giants. SIMBAD also shows a bias toward the brighter stars in our sample, thus our fainter catalogue may contain more contamination from fainter sources. In our full M dwarf candidate catalogue we find thirteen ($\sim$0.2 per cent) white dwarfs are cool enough to be selected by our initial selection process. We also find twenty-two ($\sim$0.3 per cent) of our M dwarf candidates have white dwarf companions, twenty ($\sim$0.3 per cent) are known M+M binaries, and one is a known M+L binary. 

        We also used our SIMBAD cross-match to count the source classifications given and grouped them by type (see Appendix \ref{Appendix:contamination_simbad}). From this we gauge our contamination from sources classified as galaxies, variable stars and white dwarfs as $\sim$2.7 per cent for our full M dwarf candidate catalogue. It is also interesting to note we find 1.7 per cent of our excess sample are classified as known multiple or binary systems. As with spectral type some of the source classifications carry little information (i.e. classified only as being stars or as being in an association or a cluster) therefore we also take this contamination as a rough estimate.

        We obtain additional optical spectral types by exploring data from LAMOST and we repeated this exercise with the LAMOST DR1 and DR2 catalogue spectral types (again cross-matched to three arcsec). In total there were 9,262 sources with spectral types in our full M dwarf candidate catalogue. From this we gauge our contamination from early-than-M stars and white dwarfs. The full catalogue has $\sim$9.6 per cent contamination from these sources, (see Appendix \ref{Appendix:contamination_simbad}), however it should be noted LAMOST does not distinguish between giants and dwarfs nor between spectral types of the double stars thus our contaminations are a rough estimate. In our full M dwarf candidate catalogue we find 8 ($\sim$0.1 per cent) white dwarfs are cool enough to be selected by our initial selection process.

\section{Selecting M dwarfs with mid-infrared excess}
    \label{section:selecting_Ms_with_MIR_excess}

    \subsection{Catalogue sub-sample for excess studies}
    	\label{section:Catalogue_selection:reddeneing}

        To facilitate our search for M dwarfs with mid-infrared excess we identified a sub-sample from within our M dwarf catalogue, using more stringent and additional constraints (hereinafter the `excess sample'). Our colour excess signal could be confused with interstellar reddening and/or photometric uncertainty, thus we aim to minimise their contribution. With an estimated three per cent excess from an unresolved companion (see Section \ref{section:mir_excess:excess_signatures}) we require all uncertainties to be less than this level. Reddening and photometric uncertainty cuts were designed to achieve or better this requirement, while maintaining a sufficiently high number of candidate M dwarfs.

        To enable reddening cuts we obtained extinction information from dust maps \citep{Schlegel1998}, and updated the extinctions using \citet{Schlafly2011}. We required little to no reddening, comparable to the uncertainties in the photometric data and reddening in \JWa, \JWb, \HWa, and \HWb, (i.e. $E(\JWa)$, $E(\JWb)$, $E(\HWa)$, and $E(\HWb)$  respectively) and required reddening to be less than two per cent (see Appendix \ref{Appendix:modified_reddening_equation}). After the reddening cuts 138,572 of the 440,694 M dwarfs were retained.

        To ensure high quality photometry we required that photometric magnitudes:
        \begin{enumerate}[leftmargin=*,labelindent=16pt,label=-]
         	\item had uncertainties better than 0.04 in $g$, $r$, $i$ and $z$ (413,933 sources in the full M dwarf candidate catalogue)
         	\item had uncertainties better than 0.04 in $V$, $J$, $H$, $K$, $W1$ and $W2$ (150,307 sources in the full M dwarf candidate catalogue)
            \item had unsaturated $g$ and $r$ photometry \citep[$g$>14, $r$>14, ][439,202 sources in the full M dwarf candidate catalogue]{York2000}
         	\item had WISE photometry unblended (flags $na=0$ and $nb=1$; 416,330 sources in the full M dwarf candidate catalogue)
            \item had non-variable WISE photometry \citep[see ][1,011 sources were variable in the full M dwarf candidate catalogue]{Pinfield2014}
         	\item had SDSS photometry not registering as an extended source (flag ext$\_$flg$=0$; 435,087 sources in the full M dwarf candidate catalogue)
         	\item had an SDSS score\footnote{The `score' is a number between zero and one rating the quality of an SDSS image field, see \url{http://www.sdss3.org/dr10/algorithms/resolve.php}} greater than 0.5 (407,962 sources in the full M dwarf candidate catalogue)
            \item were not flagged\footnote{We chose which SDSS flags to use by assessing the quality flags for photometric outliers in our sample. Detailed information on these flags can be found at \url{https://www.sdss3.org/dr9/algorithms/photo_flags.php}. \label{footnote:SDSS_flags}} as too close to the edge of their frames (using the {\itshape EDGE} flag; 413,944 sources in the full M dwarf candidate catalogue)
            \item were not flagged$^{\ref{footnote:SDSS_flags}}$ as using photometry from bad images (using the {\itshape PEAKCENTER}, {\itshape NOTCHECKED} and {\itshape DEBLEND\_NOPEAK}; 439,641, 429,979 and 419,436 sources respectively in the full M dwarf candidate catalogue)
            \item were not flagged$^{\ref{footnote:SDSS_flags}}$ as having photometry from images containing saturated pixels ({\itshape SATURATED}; 416,889 sources in the full M dwarf candidate catalogue)
            \item were not flagged$^{\ref{footnote:SDSS_flags}}$ as having more than 20 per cent of the point spread function flux interpolated (using the {\itshape PSF\_FLUX\_INTERP} flag; 380, 868 sources in the full M dwarf candidate catalogue)
        \end{enumerate}
        Combining all of these cuts left 36,898 M dwarf candidates in our excess sample. These cuts effectively remove the galactic plane from our excess sample (one is within galactic latitude of $\pm$15$^\circ$ and 255, 0.7 per cent, are within galactic latitude of $\pm$20$^\circ$).

    \begin{figure*}
       \begin{center}
            \includegraphics[trim=4.0cm 2.5cm 0cm 0cm, clip=true, width=0.9\textwidth]{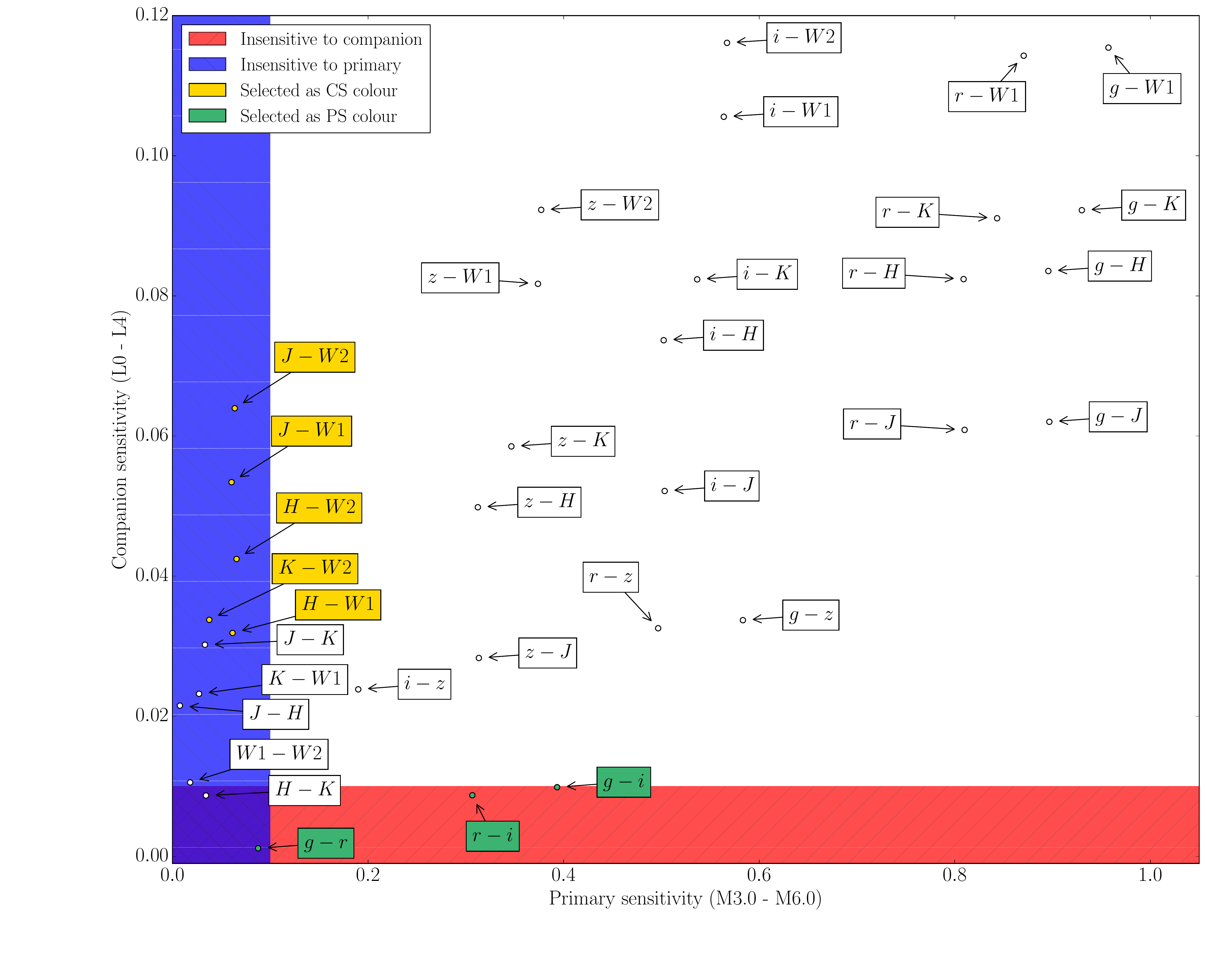}
            \end{center}
            \caption{Colour excess due to a companion (companion sensitivity) is plotted against the change in primary colour for delta-spectral-type=1.0 (primary sensitivity). The results from Appendix \protect\ref{Appendix:simulation} have been averaged for L0-L4 companions, and for M3-M6 primaries. These are expected to be the most common systems that our analysis will identify. Regions that are insensitive to companions and to spectral type variations are shaded blue and red respectively. Our chosen CS and PS colours, see text, are indicated in yellow and green respectively) \label{figure:colour_sense}.}
    \end{figure*}

	\subsection{Simulating photometry}
		\label{section:mir_excess:sim_phot}

        Although M dwarf colours are intrinsically scattered at some level\footnote{M dwarf colours may be intrinsically scattered by many factors including differences in temperature, surface gravity and composition, e.g. in \citet{Burrows1997}; rotation, e.g. in \citet{McQuillan2014}; and activity, e.g. in \citet{Robertson2013}. \label{footnote:scattered}}, the effects of adding an unresolved binary companion may be well determined. As a tool in our analysis we thus simulated M dwarf and UCD photometry which we used to interpret the observational parameter-space of the excess sample.

        For M dwarfs we constructed a probabilistic fitting routine (see Appendix \ref{Appendix:simulation}) which we applied to an M dwarf sample constructed using the following catalogues: {\itshape The  Spectroscopic Catalog of The 1,564 Brightest ($J$<9) M-dwarf Candidates in the Northern Sky\footnote{Accessed on-line at \url{http://heasarc.gsfc.nasa.gov/W3Browse/all/bnmdspecat.html}\label{footnote:Lepine13}}}\citep[selected from the {\itshape SUPERBLINK proper-motion catalogue};][]{Lepine2013}, {\itshape The Database of Ultra-cool Parallaxes\footnote{Accessed on-line at \url{https://www.cfa.harvard.edu/~tdupuy/plx/Database_of_Ultracool_Parallaxes.html} \label{footnote:dupuy12}}} \citep[from][]{Dupuy2012}, and {\itshape The Preliminary Version of the Third Catalog of Nearby Stars} \citep{Gliese1991}. We thus determined relationships between \MJ and infrared colours that led to synthetic absolute magnitudes in the $J$, $H$, $K$, $W1$, and $W2$ bands. In addition we used the synthesised colour-colour relations from  \citet{Covey2007} to generate SDSS magnitudes, making use of cubic spline fits for spectral types M0.5, M1.5, M3.5, M4.5, M5.5 \citep[See Table 3 from][]{Covey2007}. 

        For the UCDs we combined absolute magnitudes and colours from \citet{Hawley2002, Chiu2006} and \citet{Dupuy2012}, and used our probabilistic fitting routine to determine the full range of UCD optical-infrared magnitudes (see Table \ref{table:Sim_BD} for relationships not taken directly from \citealt{Dupuy2012}).

	\subsection{Choosing the optimal photometric colours}
		\label{section:mir_excess:excess_signatures}

        To optimise the photometric analysis of our excess sample we used our simulated M dwarf and UCD photometry to synthesis the expected changes in colour due to the presence of unresolved UCD companions, as well as the expected changes in colour due to spectral type variation. Our full results are shown in Appendix \ref{Appendix:simulation} (Figures \ref{figure:expectedexcess} and \ref{figure:expected_colour_sense}) with a representative plot shown in Figure \ref{figure:colour_sense}. This plot shows the colour excess due to a companion (companion sensitivity), against the change in primary colour for delta-spectral-type=1.0 (primary sensitivity). The results were averaged for L0-L4 companions and for M3-M6 primaries. Using this plot as a guide we selected two categories of colour. We defined `companion sensitive' (CS) colours as those that are sensitive to the presence of unresolved companions but are insensitive to variations in primary spectral type. We also defined `primary sensitive' (PS) colours as those that are sensitive to changes in the primary spectral type, but are insensitive to the presence of unresolved UCD companions. In addition we also considered sensitivity to metallicity when selecting PS colours (see \citealt{West2011} and \citealt{Newton2014}), even when there was little sensitivity to spectral type.

        Our final selection of CS colours are shown in yellow in Figure \ref{figure:colour_sense}. They all have primary sensitivity below 0.1 mag, and companion sensitivity above 0.03 mag. Our selected PS colours are shown in green, and all have secondary sensitivity below 0.01 mag. The (\RI) and (\GI) colours have good sensitivity to spectral type, while (\GR) is sensitive to metallicity \citep{West2011}.

	\subsection{Identifying excess using multi-colour parameter space}
		\label{section:mir_excess:multi_colour_space}

        \begin{figure*}
            \begin{center}
                \includegraphics[width=0.85\textwidth]{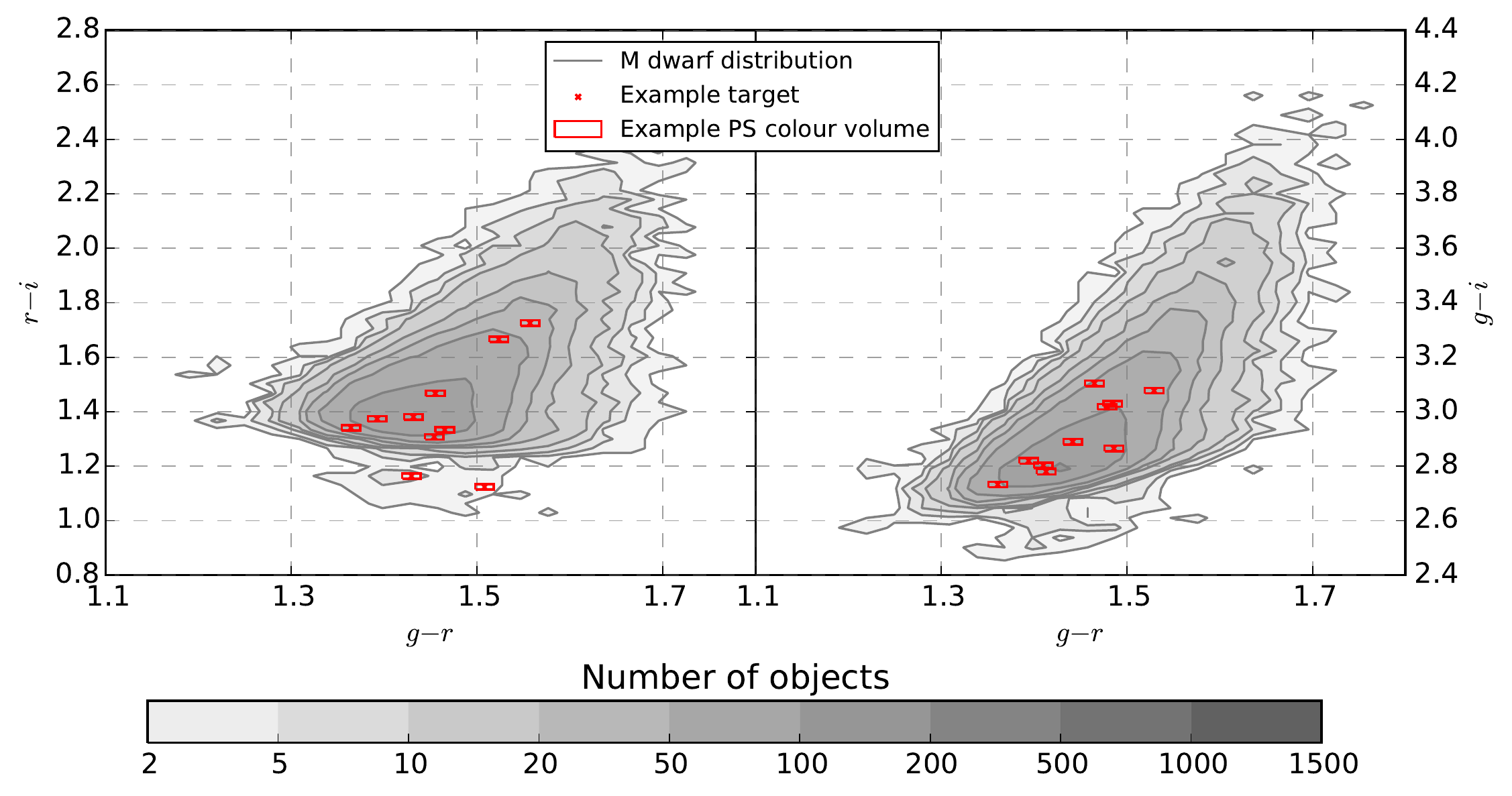}
            \caption{Colour-colour contour diagrams for \GR $\:$against \RI and, \GR $\:$against \GI, for the populations of M dwarfs used in the multi-colour-space analysis. Red boxes represent 10 targets and their respective PS colour volumes (in two dimensions). These volumes are used to define similar M dwarfs, the colour volume means and standard deviations are used to calculate colour excesses compared to those in each target colour volume. This process was undertaken for each M dwarf, except those which had less than 20 M dwarfs in their colour volume. \label{figure:colourspace}}
            \end{center}
        \end{figure*}

        \begin{figure*}
        \begin{minipage}{.475\textwidth}
                \begin{center}
                \includegraphics[height=7.5cm]{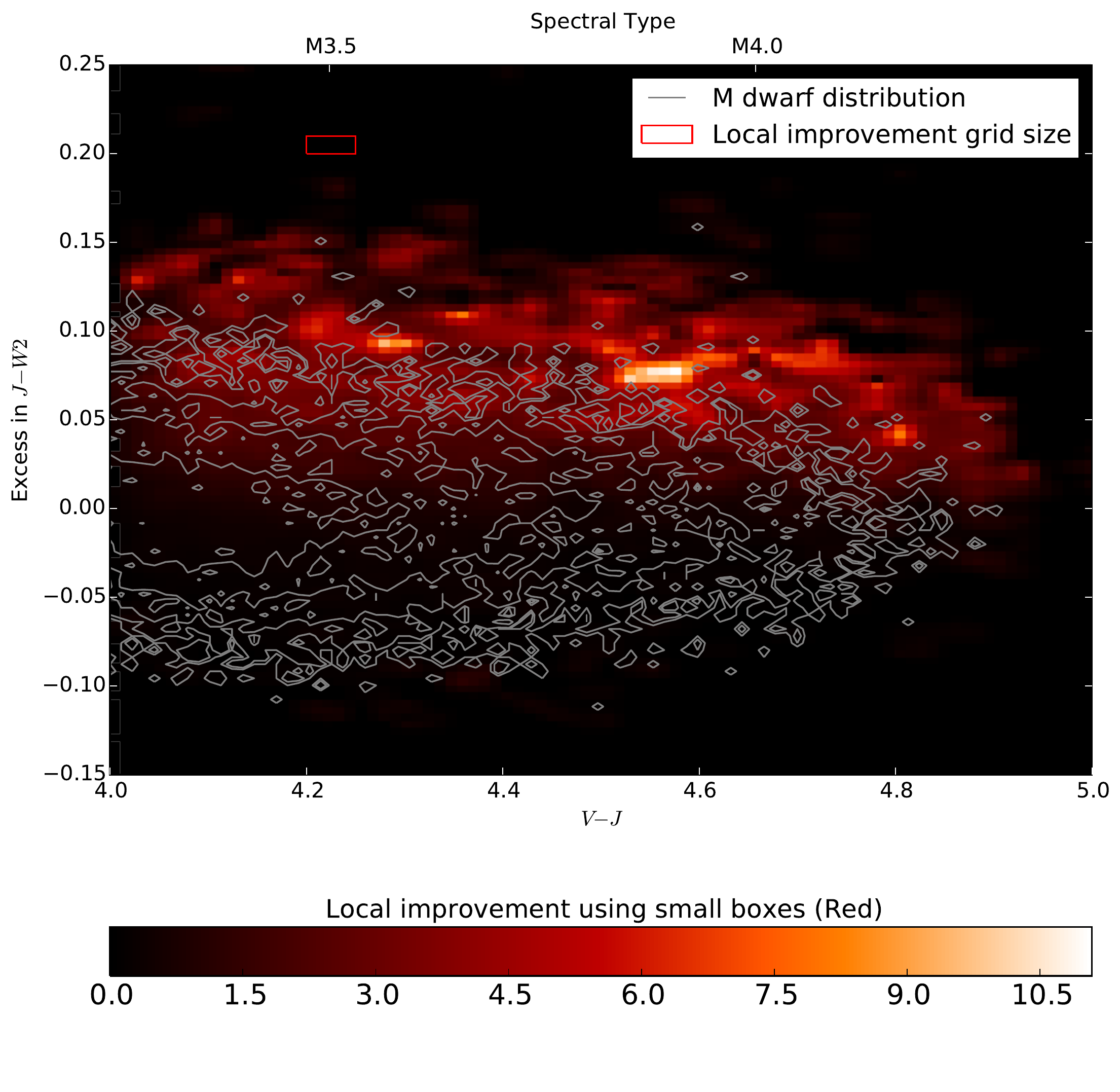}
                \end{center} 
                \caption{Excess distribution of our sample of M dwarfs after the colour-space analysis (for the CS colour \JWb). The colour distribution represents the local improvement grid created simulating M dwarf+UCD unresolved binary systems through the multi-colour-space analysis. The red box shows the size of the improvement grid used.\label{figure:improvementa}} 
        \end{minipage}\qquad
        \begin{minipage}{.475\textwidth}
            \vspace{0.75cm}
            \includegraphics[height=7.5cm]{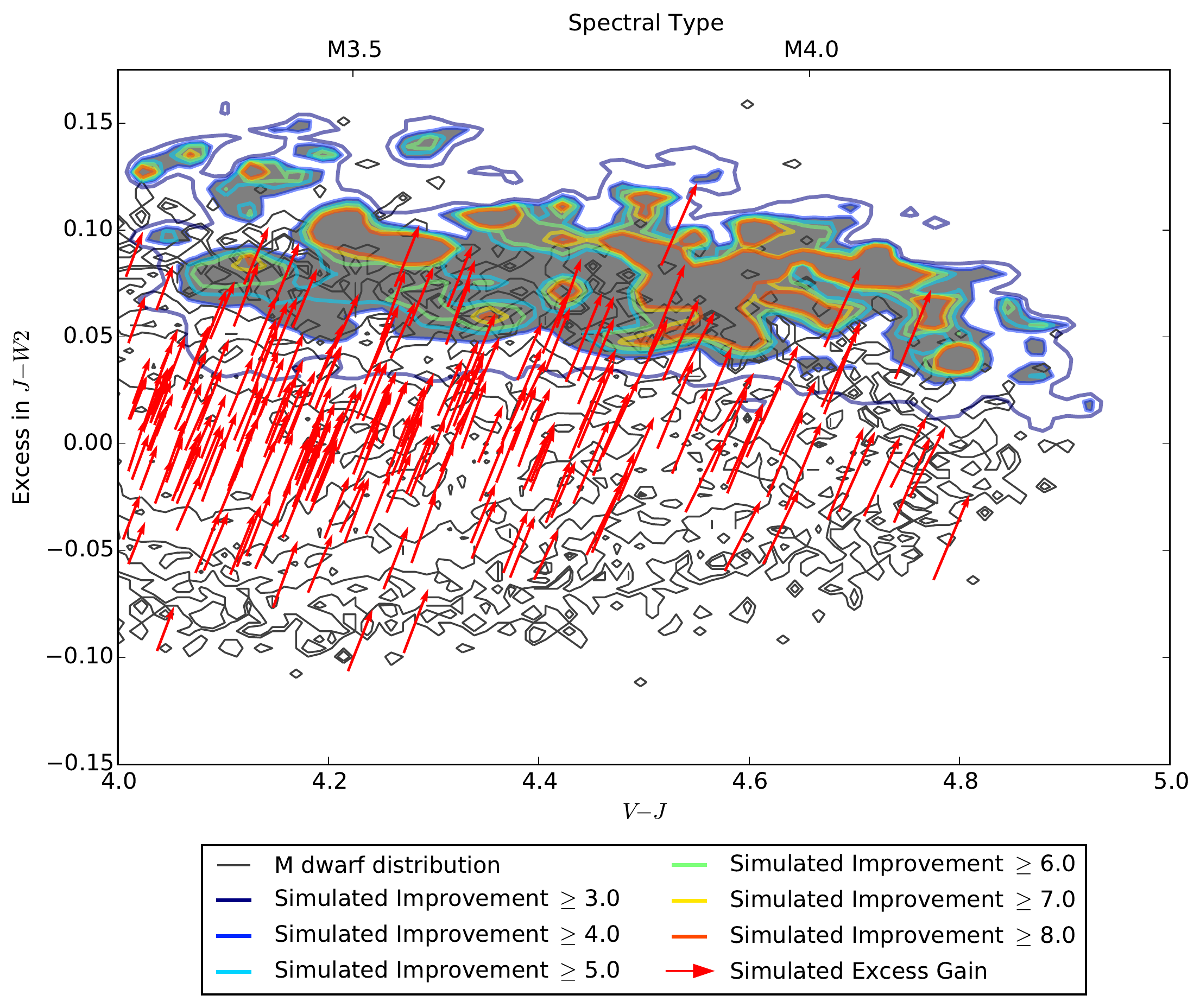}
            \caption{The local improvement grid was used to define contours of improvement (coloured contours). For comparison the excess gained (in \JWb) by a sample of M dwarfs is shown. The red arrows start at an M dwarfs location in excess$/\VJ$ space without a simulated companion and end at the location where the M dwarf + UCD companion is in excess$/\VJ$ space. The shaded region is the final selected contour equvialent to an improvement of $\ge$4 over randomly selecting an M dwarf from our full catalogue of M dwarf candidates\label{figure:improvementb}}
        \end{minipage}
        \end{figure*}

        In order to estimate the mid-infrared excess of the candidate M dwarfs in our excess sample we defined a three-dimensional colour parameter-space using the chosen PS colours (\GR), (\RI) and (\GI). For each candidate M dwarf (target M dwarf) in our excess sample we then defined a sub-volume within this PS colour-space, centred on the target M dwarf colours and with a size of $\pm$0.01 in each colour (see Figure \ref{figure:colourspace}). We then established `no companion' comparison colours for each candidate by selecting all excess sample members within a target M dwarf's PS colour sub-volume, and measured the mean CS colours in this volume. This approach assumes that the vast majority of the excess sample are M dwarfs without UCD companions, and thus the `no companion' comparison colours should provide a good zero excess reference from which the mid-infrared excess of target M dwarfs can be estimated.

        We required at least 20 comparison objects in a target M dwarf's PS colour sub-volume (this was the case for 22,579 members of the excess sample), and measured the mid-infrared excess using the most sensitive of our CS colours (\JWb). The resulting excess distribution is shown in Figures \ref{figure:improvementa} and \ref{figure:improvementb}, against \VJ (a proxy for spectral type).

        The excess distribution will be discussed further in Section \ref{section:mir_excess:Colour_excess_distribution}, and Figure \ref{figure:improvementb} also shows the selection contours that will be discussed in Section \ref{section:mir_excess:optimisation}. The excess distribution of the sample lies generally in the range -0.15 to +0.15, and as we will see (Section \ref{section:mir_excess:optimisation}) the excess values of M dwarfs with L dwarf companions lie at the upper end of this range (see also the L dwarf excess vectors shown in Figure \ref{figure:improvementb} as a guide). We note that these are significantly lower excess levels than have been previously analysed in the context of M dwarf disc excess. Several studies \citep{Esplin2014, Theissen2014, Luhman2012b} have identified M dwarfs that may have discs, by selecting those with mid-infrared excess values of $\sim$1 or greater. These studies probe excess levels that are $\sim$5 times greater than we focus on here, since M dwarf discs are generally much brighter in the mid-infrared than UCDs.

    \subsection{Colour excess distribution}
        \label{section:mir_excess:Colour_excess_distribution}

        \begin{figure*}
            \begin{center}
            \includegraphics[width=\textwidth]{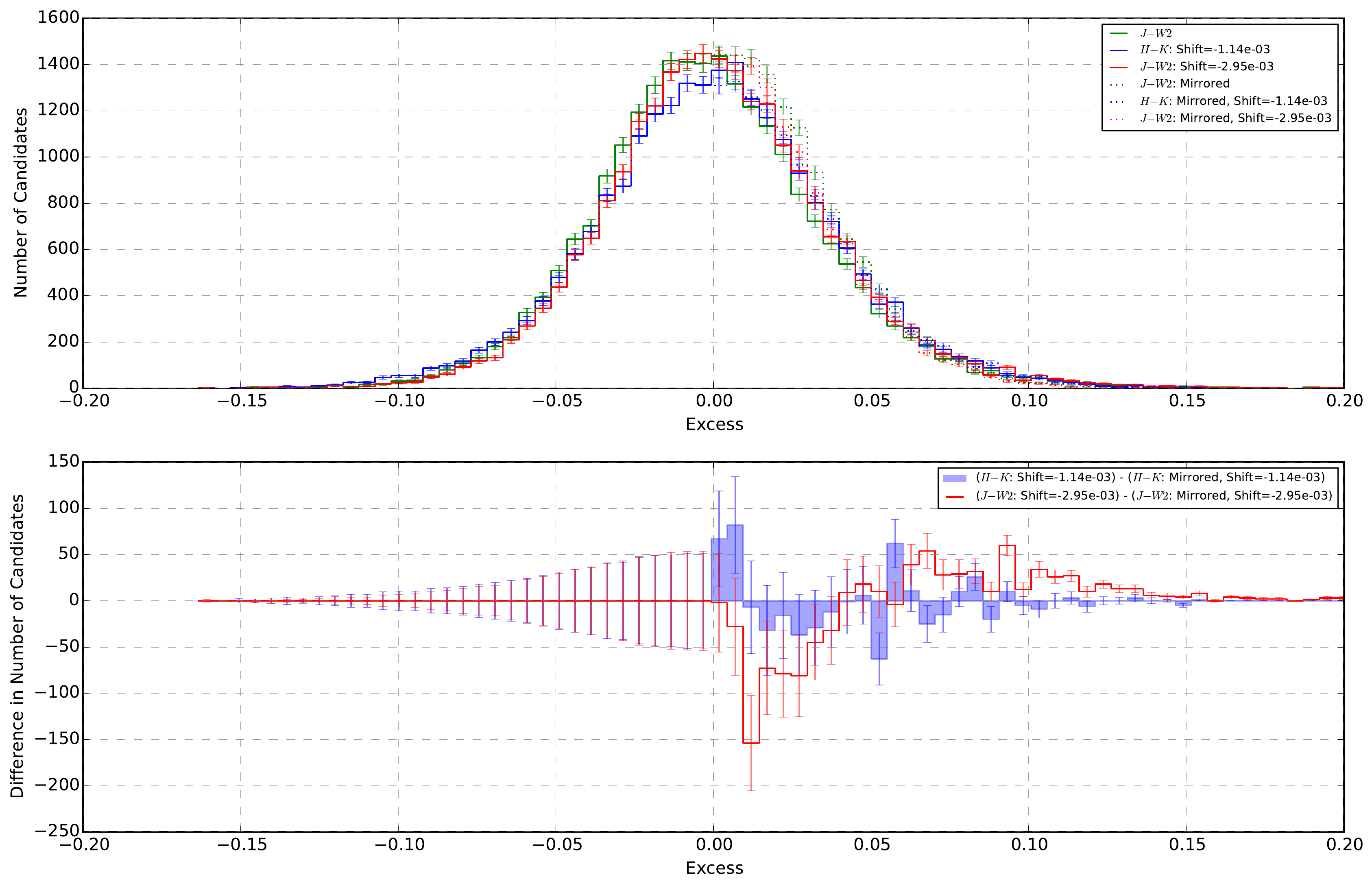}
            \caption{Assuming the deficit section of the excess \JWb distribution is from Gaussian, random or likewise symmetric contamination process (and not from a UCD companion), we show the histogram of our excess distribution, compared to a mirrored deficit (negative) distribution around zero (non-shifted, magenta) and around an overall deficit value of -0.0030 (shifted, red). This is then compared to the excess \HK distribution (an independent colour that also shows a very low sensitivity to companions and spectral type, see Figure \ref{figure:colour_sense}, shown in cyan and blue for non-shifted and shifted respectively). The residuals of \JWb are also shown in red (bottom plot). The positive bump at an excess of >0.05 magnitudes is evidence of processes contributing to making the targets redder than their mean PS colour volume colour, including M dwarfs with unresolved UCD companions and is not seen in our companion insensitive (\HK) residuals in blue. \label{figure:neg_mirror_jw2}}
            \end{center}
        \end{figure*}

        Figure \ref{figure:neg_mirror_jw2} (top plot) shows the histogram of our excess measurements. Overall it is similar to a Gaussian distribution, but as we will see it has some asymmetries. Firstly it is apparent that the peak of this histogram is found at a slightly negative excess value which seems unexpected. However, we believe this bias is introduced by our analysis method, and results from the finite size of the PS colour sub-volumes. The number density of objects varies across the sub-volumes in our multi-colour parameter-space, leading to average values that can be slightly different to the central value. We suggest that on average this effect leads to the small negative offset that is seen. For our symmetry analysis we offset the histogram by +0.003 magnitudes to remove this offset.

        To assess the symmetry of the histogram we reflected the negative side of the distribution in the Y-axis and subtracted this from the positive distribution (see Figure \ref{figure:neg_mirror_jw2}). Although the histogram is fairly symmetrical, it contains an important feature. The positive wing has relatively lower frequencies (compared to the negative wing) for excesses of 0-0.05, and has relatively higher frequencies for excesses of 0.05-0.15 (an excess bump). To assess the nature of this excess bump we carried out a comparison analysis using the \HK colour as our CS colour (instead of \JWb). This comparison analysis should not be sensitive to companion excesses, or indeed to spectral type variations (see Figure \ref{figure:colour_sense}), but should produce the kind of distribution we expect in the absence of any significant excess (albeit with some scatter due to a metallicity spread). Figure \ref{figure:neg_mirror_jw2} also contains the histogram for the \HK excess distribution (green line), and it can be see that it has a similar form to the \JWb excess histogram. However, when one studies the symmetry of this histogram (bottom plot) it is clear that the \HK excess distribution is much more symmetrical by comparison. The mirror-subtracted trace for the \HK excesses is close to zero with just a few short-range deviations. This contrasts with the bump feature seen when excess values are calculated using \JWb, and thus supports the idea that the bump is caused by a population of M dwarfs with mid-infrared excess, rather than by some unidentified bias in our analysis method.

        The mid-IR bump represents $\sim$2.01 to $\gtrsim$2 per cent of our excess sample, and we thus expect unresolved M+UCD systems to only form a fraction of this population (see further treatment in Section \ref{section:mir_excess:optimisation}). We also expect the bump population to include a variety of contaminating objects such as M+M binaries (where the cooler companion causes an excess), M dwarfs in regions of local reddening (not picked up by our reddening assessments), and M dwarfs with some low level of disc emission. These objects will be mixed with M dwarfs whose colours have scattered to the red due to photometric uncertainty.

    \subsection{Excess selection contours}
    	\label{section:mir_excess:optimisation}
       
        In order to identify M dwarfs likely to have mid-infrared excess consistent with unresolved UCD companions, we used our simulated M dwarf and UCD photometry (from Section \ref{section:mir_excess:sim_phot}). As a starting point we took the photometry of our excess sample to represent a population without any unresolved UCD companions. This assumes that UCD companions are reasonably rare, which is consistent with previous constraints (Section \ref{section:intro})) and our interpretation of the excess bump feature. We then simulated unresolved UCD companions around a randomly selected fraction ($\beta$) of our sample by modifying the M dwarf colours to account for L2 companions (since we expect the most significant UCD reddening from companions in the range $\sim$L0-L3; see Appendix \ref{Appendix:simulation}). We used these simulated M dwarfs to map out a so-called `improvement' parameter-space. We define `improvement' to be the factor by which the probability improves that an M dwarf has an unresolved UCD companion, compared to a completely random selection.

        \begin{equation}
            Improvement = \frac{N_{SB}}{N_T}\cdot\frac{1}{\beta}
            \label{equation:mir_excess:Improvement}
        \end{equation}
        \\
        where $N_{SB}$ is the total number of simulated M dwarf+UCD unresolved binary systems present in the region, $N_T$ is the total number of M dwarfs present in a region ($N_T = N_D - N_{SS} + N_{SB}$), and $\beta$ is the simulated binary fraction. Here $N_D$ is the number of original M dwarfs in the region and $N_{SS}$ is the number of simulated M dwarf+UCD unresolved binary systems present in the region before the UCDs were added. Hence an improvement of one is equivalent to no improvement (i.e. randomly selecting M dwarfs from the distribution).

        We calculated improvement values across the excess \VJ parameter-space of our excess sample using a box-smoothed approach and running our simulation 1000 times to smooth out the random noise. Figure \ref{figure:improvementa} shows the `improvement' levels (colour-scaled) across the excess \VJ parameter-space. The box size we used for smoothing is indicated in the upper left of the diagram.

        A set of improvement contours were defined to aid selection of potential M+UCD binaries. These are shown in Figure \ref{figure:improvementb}, where the contours range from 3-8. We required improvement $\ge$4 for our final selection, and this region is shaded in grey in the Figure. We used these contours as selection regions for our candidates and the results can be seen in Table \ref{table:Improvement_From_Contours}, for a simulated binary fraction of 0.01 (1 per cent). For \JWb and $\beta$=0.01 this led to 1,082 objects which constitutes our `candidate M+UCD sample'.

        \begin{table*}
           \begin{center}
            \begin{tabular}{ccccccc}
            \hline
            Colour & $Imp\ge$ 3 & $Imp\ge$ 4 & $Imp\ge$ 5 & $Imp\ge$ 6 & $Imp\ge$ 7 & $Imp\ge$ 8 \\
            \hline
            \JWa & 1,800 & 654 & 330 & 169 & 110 & 83 \\
            \JWb & 2,934 & 1,082 & 511 & 269 & 128 & 82 \\
            \HWa & 705 & 176 & 85 & 34 & 26 & 15 \\
            \HWb & 1,095 & 301 & 118 & 57 & 23 & 17 \\
            \KWb & 616 & 221 & 98 & 34 & 14 & 8 \\
            \hline
            \end{tabular}
            \end{center}
            \caption{Using a locally defined Improvement, $Imp$, based on our simulated M dwarf+UCD unresolved binary systems, contours of improvement were defined (i.e. $Imp\ge$ 3) and the number of M dwarfs in excess/\VJ space which fell inside a contour were recorded. The simulated binary fraction, $\beta$ in this case was $\beta=0.01$ (1 per cent). \label{table:Improvement_From_Contours}}      
        \end{table*}

    \subsection{Measuring improvement in detection}
        \label{section:mir_excess:improvement}
        
\begin{figure*}
        \begin{minipage}{.45\textwidth}
            \begin{center}
            \vspace{0.75cm}
            \includegraphics[height=8cm]{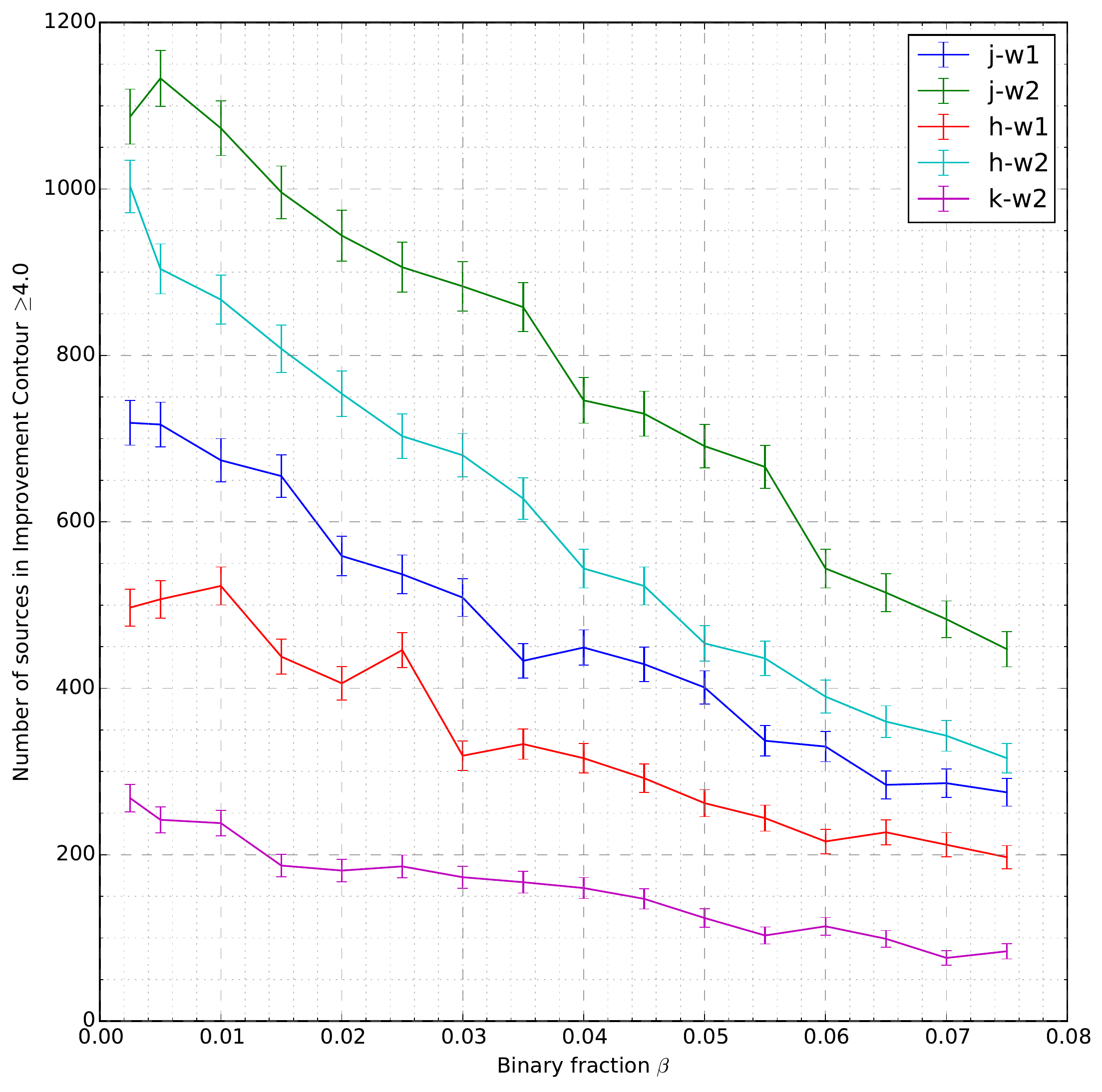}
            \end{center}
            \caption{After varying the initial binary fraction of our simulated companion systems, the final detected fraction was calculated and thus the yield of candidate M+UCDs was calculated and the improvement over randomly selecting M dwarf was defined for each CS colour. Plot shows candidate M+UCDs for an improvement $\ge$ 4. \JWb yields the most candidate M+L dwarf systems.\label{figure:num_vs_imp}}
        \end{minipage}\qquad
        \begin{minipage}{.45\textwidth}
            \begin{center}
                \includegraphics[height=8cm]{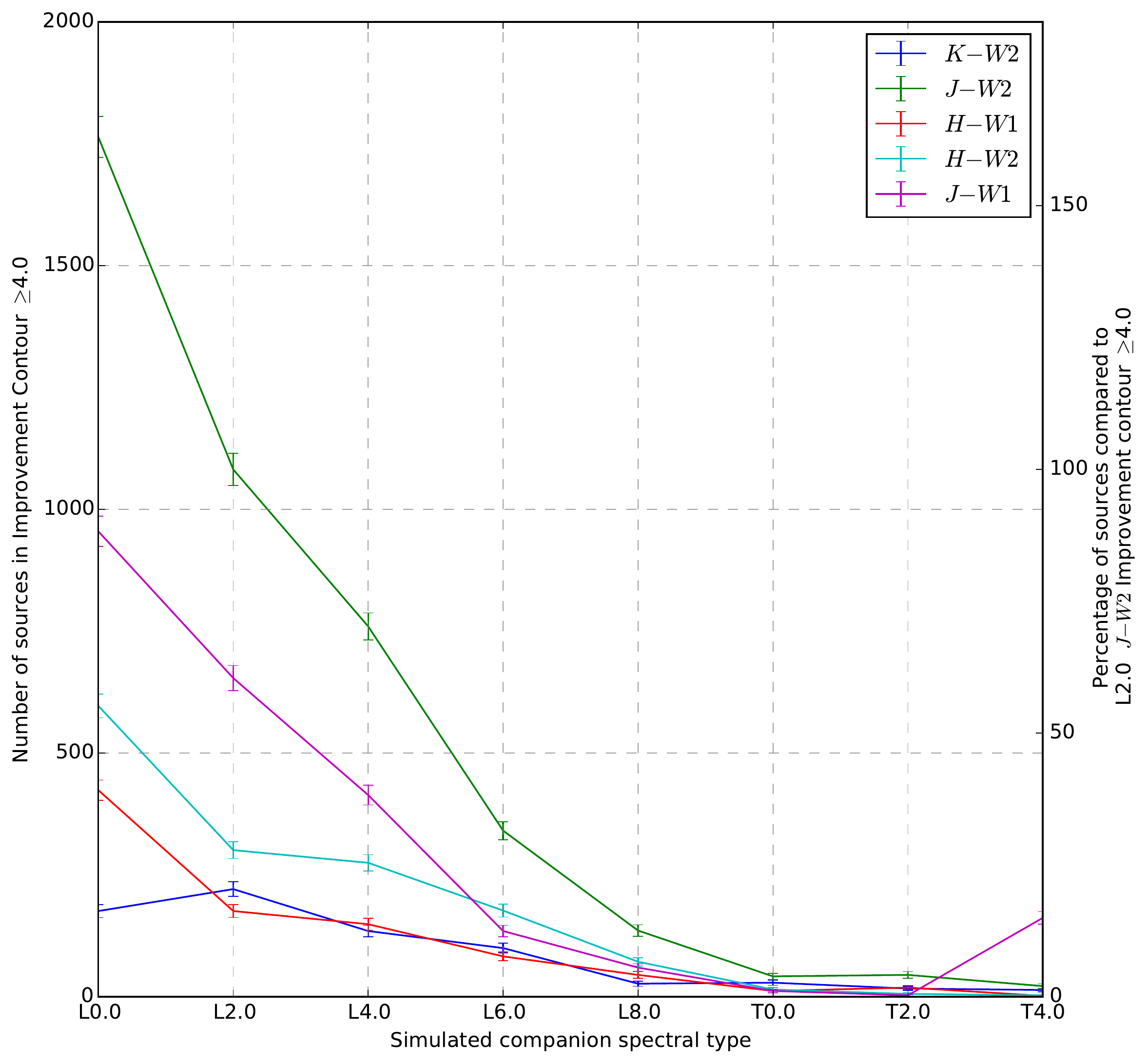}
            \end{center}    
            \caption{After varying the simulated companion between L0 and T4 we are able to gain insight into the possible companion spectral type distribution. As with Figure \ref{figure:num_vs_imp}, \JWb yields the most candidate M+L dwarf systems. \label{figure:predicted_subtype}}
        \end{minipage}
        \end{figure*}

        By varying the fraction of simulated binaries added ($\beta$, Section \ref{section:mir_excess:optimisation}) we were able to estimate the expected yield of candidate M+UCDs at certain improvement levels. The binary fraction for mid-type M dwarfs with a UCD companion is rather uncertain, so we present a range of estimates for $\beta$=0.2-8 per cent. We run the same selection method as above to create additional candidate M+UCD samples, with the only difference being $\beta$. We count the number of candidates found for each binary fraction and show this in Figure \ref{figure:num_vs_imp}. The higher the binary fraction the lower our yield, this is expected because more of our reference PS colour M dwarfs have companions thus diluting the colour excess detectable. For a binary fraction of 0.01 we expect over 1000 candidate M+UCDs for \JWb.

    \subsection{Predicting candidate companion subtype}
        \label{section:mir_excess:prediction}

        To estimate the subtypes of the expected companions we ran our simulation (Section \ref{section:mir_excess:optimisation}) for L0-T4 companions (in steps of two subtypes). We run the same selection method as above (except varying the companion we added) to create additional candidate M+UCD samples. Figure \ref{figure:predicted_subtype} shows the result for the number of candidate M+UCD systems with a binary fraction of 0.01. For comparison we show the predictions when a range of other CS colours are used instead of \JWb. It can be seen that \JWb yields the most candidate M+L dwarf systems because it is the most sensitive to M+L unresolved binaries. If the companion distribution is flat, similar to the field population \citep[See Figures 11 and 12 from][]{Cruz2003} we can use Figure \ref{figure:predicted_subtype} to predict roughly our expectation of companion subtypes. We expect up to 60 per cent of our candidates to have companions of spectral subtype earlier than L3 and $\sim$35 per cent to be later L dwarfs. The remaining $\sim$5 per cent may be late L dwarfs or early T dwarfs.

    \subsection{Contamination in the excess sample and candidate M+UCDs}
        \label{section:mir_excess:contamination}

        In total there were 3,928 matches out of the 36,898 excess sample and 66 matches out of the 1,082 candidate M+UCDs to the SIMBAD catalogue (three arcsec cross-match). Of these 1,475 and 32 respectively had spectral types from SIMBAD. From this we gauge our contamination from early (FGK) stars, M giants, and white dwarfs. Our excess sample has a contamination of $\sim$0.14 per cent from these sources and there is no contamination from these sources found in our candidate M+UCDs (see Table \ref{table:SIMBAD_spt_classifications}). It should however be noted, as with the full catalogue, some of the spectral types are defined only as M type star ($\sim$1.35 per cent) and thus we may slightly underestimate our contamination from M giants. 

        We repeated this exercise with the LAMOST DR1 and DR2 catalogue spectral types (again with a three arcsec cross-match). In total there were 1,851 with spectral types out of our 36,898 excess sample and 41 with spectral types out of our 1,082 candidate M+UCDs. From this we gauge our contamination from early stars, multiple stars (which we assume are all from contamination) and white dwarfs. Our excess sample has a contamination of $\sim$2.38 per cent from these sources and there is a $\sim$2.44 per cent contamination from these sources found in our candidate M+UCDs (see Table \ref{table:LAMOST_spt_classifications}). However, as with the full catalogue, it should be noted LAMOST does not distinguish between giants and dwarfs nor between spectral types of the double stars thus our contaminations are a rough estimate.

        We counted the source classifications given in SIMBAD and grouped them (see Table \ref{table:SIMBAD_type_classifications}). From this we gauge our contamination from galaxies, variable stars and white dwarfs as $\sim$0.97 per cent for our excess sample and $\sim$1.52 per cent for our candidate M+UCDs. As with the spectral types some of the source classifications are not specific enough to gauge possible contamination (i.e. classified as stars or as being in an association or a cluster) therefore we take these contaminations as rough estimates.

\section{Summary and future work}
    \label{section:summary}
    We present a new photometric method to target unresolved UCD companions to M dwarfs. Using WISE, 2MASS and SDSS we create a large catalogue of 440,694 M dwarfs candidates. By requiring high accuracy photometry in un-reddened regions of sky, we isolate a sample of 36,898 catalogue members and search for \JWb outliers in tightly constrained multi-colour (\GR, \GI, and \RI) sub-volumes centred on each object. These colours were chosen to optimise our methodology, which isolates a comparison sample of M dwarfs (similar $T_{eff}$ and metallicity) for each target M dwarf, and then measures mid-infrared excess with respect to this comparison. 

    We select a region in excess \VJ parameter-space where the likelihood of systems is greater than a factor of four over the chance of randomly selecting a companion, assuming a companion fraction of $\sim$0.01. In total we obtain 1,082 candidate M+UCDs for \JWb. We discuss the excess distribution and conclude there is good evidence for an overall excess in \JWb in our distribution due to unresolved UCD companions. Based on simulation we expect up to 60 per cent of our candidates to have companions of spectral subtype earlier than L3 and $\sim$35 per cent to be later L dwarfs. The remaining $\sim$5 per cent may be late L dwarfs or early T dwarfs. For our full catalogue, excess sample and candidate M+UCDs we estimate the contamination using SIMBAD and LAMOST ($\sim$3 per cent), thus confirming that we have very high quality, clean samples of M dwarfs.

    Further analysis of our candidate M+UCD binary sample is needed to confirm the UCD companions. Optical spectroscopy will confirm M dwarfs and identify any whose colours are scattered away from their typical location (thus leading to contamination in the selection). High resolution imaging (e.g. adaptive optics, lucky imaging, HST) may reveal the UCD companions directly if their angular separation is $\gtrsim$0.1 arcsec. Closer systems may be constrained through radial velocity measurements, with very close systems potentially being amenable to transit light curve studies. Gaia may also be capable of constraining some systems if there is a detectable astrometric wobble.

\section*{Acknowledgements}

    NJC acknowledges support from the UK's Science and Technology Facilities Council [grant number ST/K502029/1], and has benefited from IPERCOOL, grant number 247593 within the Marie Curie 7th European Community Framework Programme. MG acknowledges support from Joined Committee ESO and Government of Chile 2014 and Fondecyt Regular No. 1120601. Support for MG and RGK is provided by the Ministry for the Economy, Development, and Tourisms Programa Inicativa Cientifica Milenio through grant IC 12009, awarded to  The  Millennium  Institute  of  Astrophysics (MAS) and acknowledgement to CONICYT REDES No. 140042 project. R.G.K. is supported by Fondecyt Regular No. 1130140.  We make use of data products from WISE \citep{Wright2010}, which is a joint project of the UCLA, and the JPL$/$CIT, funded by NASA, and 2MASS \citep{Skrutskie2006}, which is a joint project of the University of Massachusetts and the Infrared Processing and Analysis Center$/$CIT, funded by NASA and the NSF. We also make substantial use of SDSS DR10, funding for SDSS-III has been provided by the Alfred P. Sloan Foundation, the Participating Institutions, the NSF, and the USDOESC. This research has made use of the NASA$/$IPAC Infrared Science Archive$^{\ref{footnote:IRSA}}$, which is operated by JPL, CIT, under contract with NASA, and the VizieR database catalogue access tool and SIMBAD database\citet{Wenger2000}, operated at CDS, Strasbourg, France. This work is based in part on services provided by the GAVO Data Center and the data products from the PPMXL database of \citet{Roeser2010}. This publication has made use of LAMOST DR1 and DR2 spectra. Guoshoujing Telescope (LAMOST) is a National Major Scientific Project built by CAS. Funding for the project has been provided by the National Development and Reform Commission. LAMOST is operated and managed by the NAO, CAS. This research has benefited from the SpeX Prism Spectral Libraries, maintained by Adam Burgasser$^{\ref{footnote:Spex}}$. This research made extensive use of: {\sc Astropy} \citep{Astropy2013}; {\sc matplotlib} \citep{Chabrier2007}, {\sc scipy} \citep{jones2001}; {\sc Topcat} \citep{Topcat}; {\sc Stilts} \citep{Stilts} {\sc ipython} \citep{Perez2007} and NASA's Astrophysics Data System.

\bibliographystyle{mnras}

\appendix

\section{Estimates on contamination}
    \label{Appendix:contamination_simbad}

    \counterwithin{figure}{section}

    \begin{table*}
       \begin{center}
        \begin{tabular}{p{1.5cm}p{9.5cm}p{1.5cm}p{1.5cm}p{1.5cm}}
        \hline
        Group & SIMBAD spectral type selected for group & Number in full candidate catalogue & Number in excess sample & Number in candidate M+UCDs \\ 
        \hline
        Total & - & 440,694 & 36,898 & 1,082 \\
        Total (with SIMBAD) & - & 7,360 & 1,475 & 32 \\
        \hline
        White dwarf & {\itshape DA, DA.7, DA1.1, DA1.7, DA2.9, DA3, DA3.3, DA3.5, DB, DC..., DC-DQ} & 13 (0.18\%) & 1 (0.07\%) & 0 \\
        White dwarf binaries & {\itshape D+M, DAM, DA+M, DA+dM, DA+dM:, DA+dMe, DA+M3V, DA+M4, DB+..., DB+M, DB+M3 DO+M, DC+M, DC+dM} & 22 (0.30\%) & 0 & 0 \\ 
        F         & {\itshape F9.5} & 1 (0.01\%) & 0 & 0 \\
        G         & {\itshape G:, G2III} & 2 (0.03\%) & 0 & 0 \\
        K         & {\itshape K, K:, K..., K/M} & 15 (0.20\%) & 1 (0.07\%) & 0 \\
        early K   & {\itshape K3, K4, K4.5, K4/5} & 9 (0.12\%) & 0 & 0 \\
        late K    & {\itshape K4V:, K5, K5V, K5Ve K5.3, K5/M0, K6, K6V, K6Ve, K6.5, K7, K7V, K8, K9V} & 56 (0.76\%) & 0 & 0 \\
        M         & {\itshape M, M:, MV:, MV, MV:e} & 145 (1.97\%) & 36 (2.44\%) & 0 \\
        M0 - <M1  & {\itshape M0V:, M0Vk, M0, M0V, M0e, M0.4, M0.5, M0.5V, M0.6, M0.8} & 38 (0.52\%) & 2 (0.14\%) & 0 \\
        M1 - <M2  & {\itshape M1V, M1, M1.0, M1.0V, M1e, M1.5, M1.5V} & 48 (0.65\%) & 1 (0.07\%) & 0 \\
        M2 - <M3  & {\itshape M2, M2.0, M2V, M2.0V, M2e, M2V:, M2.3, M2.4, M2.4V, M2.5, M2.5V, M2.6, M2.7, M2.8, M2.9, M2/3} & 169 (2.30\%) & 24 (1.63\%) & 0 \\
        M3 - <M4  & {\itshape M3.0, M3, M3e, M3V, M3V:, M3.0V, M3.1, M3.2, M3.3, M3.3V, M3.4, M3.5, M3.5V, M3.5e, M3.6, M3.7, M3.8, M3.9, M3..., M3:, M3-4} & 1099 (14.93\%) & 159 (10.78\%) & 7 (21.86\%) \\
        M4 - <M5  & {\itshape M4V, M4.0V, M4, M4.0, M4.1, M4.2, M4.25V, M4.3, M4.3V, M4.4, M4.4V, M4.5, M4.5V, M4.6, M4.6V, M4.7, M4.7v..., M4.75, M4.75V, M4.8, M4.9, M4-5, M4..., M4:V} & 2663 (36.18\%) & 642 (43.53\%) & 13 (40.63\%) \\
        M5 - <M6  & {\itshape M5, M5e, M5V, M5.0, M5.0V, M5V:, M5Ve, M5.1, M5.2, M5.2, M5.3, M5.4, M5.4V, M5.5, M5.5V, M5.7, M5.9, M5.9V, M5..., M5V:e...} & 1189 (16.15\%) & 330 (22.37\%) & 12 (37.5\%) \\
        M6 - <M7  & {\itshape M6, M6.0, M6.0V, M6e, M6V, sdM6, M6-M6.25, M6.1, M6.2v..., M6.3, M6.4, M6.5, M6.5V, M6e...} & 1053 (14.31\%) & 173 (11.73\%) & 0 \\
        M7 - <M8  & {\itshape M7.0, M7, M7V, M7.0V, M7.5} & 735 (9.99\%) & 100 (6.78\%) & 0 \\
        M8 - <M9  & {\itshape M8, M8V} & 74 (1.01\%) & 1 (0.07\%) & 0 \\
        >M9       & {\itshape M9V} & 4 ( 0.05\%) & 0 & 0 \\
        early L   & {\itshape L0, L1.5} & 2 (0.03\%) & 0 & 0 \\ 
        M giants  & {\itshape M3III} & 1 (0.01\%) & 0 & 0 \\
        M + M binaries & {\itshape M0+M1, M2+M3, M2+M5, M2.5+M3.5, M2.5+M4.0, M3+M3, M3+M4, M3.5+M4.0, M3+WD, M4+M4, M4+WD, M4.2+M4.3, M4.5+M5.5, M5.0+M6.0, M6+WD} & 20 (0.27\%) & 5 (0.07\%) & 0 \\
        M + L binaries & {\itshape M80v+L3.0V} & 1 (0.01\%) & 0 & 0 \\
        \hline
        Non contaminated sources & {\itshape M, M0 - <M1 to M9> early L, D+M, M+M binaries, M+L binaries} & 7,263  (98.68\%) & 1,473 (99.86\%) & 32 (100.00\%) \\
        Contaminated sources & {\itshape D, F, G, K, early K, late K, M3 Giants} & 97 (1.32\%) & 2 (0.14\%) & 0 \\
        \hline
        \end{tabular}
        \end{center}
        \caption{Statistics on SIMBAD spectral types for the cross-match between the full catalogue of M dwarf candidates, the excess sample and the candidate M+UCDs with SIMBAD. Spectral types are only shown for those SIMBAD spectral types with non-zero cross-matches. Note some spectral types have no subtype, e.g. `M', and thus for these sources we cannot identify whether they are dwarfs or giants (and thus whether these sources contribute to the contamination) \label{table:SIMBAD_spt_classifications}}
    \end{table*}

    \begin{table*}
       \begin{center}
        \begin{tabular}{p{4cm}p{5cm}p{3cm}p{2cm}p{2cm}}
        \hline
        Group & LAMOST spectral types selected for group & Number in full candidate catalogue & Number in excess sample & Number in candidate M+UCDs \\
        \hline
        Total & - & 440,694 & 36,898 & 1,082 \\
        Total with LAMOST spectral types & - & 9,262 & 1,851 & 41 \\
        \hline
        A & {\itshape A0, A1IV, A1V, A2V, A4III, A6V, A7IV} &  8 (0.09\%) & 0 & 0 \\
        D & {\itshape WD, WDMagnetic} & 8 (0.09\%) & 1 (0.05\%) & 0 \\
        F & {\itshape F0 F2 F3 F4 F5 F6 F7 F9} & 42 (0.45\%) & 6 (0.32\%) & 0 \\
        G & {\itshape G0 G1 G2 G3 G4 G5 G6 G7 G8 G9} & 286 (3.09\%) & 33 ( 1.78\%) & 1 (2.44\%) \\ 
        early K & {\itshape K0 K1 K2 K3 K4} & 44 (0.48\%) & 1 (0.05\%) & 0 \\
        late K & {\itshape K5 K7} 503 (5.43\%) & 3 (0.16\%) & 0 \\
        M0 - <M1  & {\itshape M0 M0V} & 540 (5.83\%) & 2 (0.11\%) & 0 \\
        M1 - <M2  & {\itshape M1} & 442 (4.77\%) & 2 (0.11\%) & 0 \\ 
        M2 - <M3  & {\itshape M2 M2V} & 827 (8.93\%) & 110 (5.94\%) & 2 (4.88\%) \\
        M3 - <M4  & {\itshape M3} & 5,874 (63.42\%) & 1,505 (81.31\%) & 33 (80.49\%) \\
        M4 - <M5  & {\itshape M4} & 607 (6.55\%) & 154 (8.32\%) & 5 (12.20\%) \\
        M5 - <M6  & {\itshape M5} & 11 (0.12\%) & 2 (0.11\%) & 0 \\
        M6 - <M7  & {\itshape M6} & 28 (0.30\%) & 8 (0.43\%) & 0 \\
        M7 - <M8  & {\itshape M7} & 0 & 0 & 0 \\
        M8 - <M9  & {\itshape M8} & 0 & 0 & 0 \\
        >M9       & {\itshape M9} & 2 (0.02\%) & 0 & 0 \\
        double star & {\itshape DoubleStar} & 40 (0.43\%) & 7 (0.38\%) & 0 \\
        \hline
        Non contaminated sources &  double star, M0 - <M1 to M9>, early L & 8,371 (90.38\%) & 1807 (97.62\%) & 40 (97.56\%) \\
        Contaminated sources & D, A, F, G, early K, late K & 891 (9.62\%) & 44 (2.38\%) & 1 (2.44\%) \\
        \hline
        \end{tabular}
        \end{center}
        \caption{Statistics on LAMOST source classifications for the cross-match between the full catalogue of M dwarf candidates, the excess sample and the candidate M+UCDs with LAMOST. Object classifications are only shown for those LAMOST spectral types with non-zero cross-matches. Note spectral types are only given to integer spectral types and giants and dwarfs are not distinguished. \label{table:LAMOST_spt_classifications}}
    \end{table*}

    \begin{table*}
       \begin{center}
        \begin{tabular}{p{4cm}p{6cm}p{2cm}p{2cm}p{2cm}}
        \hline
        Group & SIMBAD Object Types selected for group & Number in full candidate catalogue & Number in excess sample & Number in candidate M+UCDs \\
        \hline
        Total & - & 440,694 & 36,898 & 1,082 \\
        Total with SIMBAD cross-matches & - & 20,286 & 3,928 & 66 \\
        \hline
        Potential M dwarfs & {\itshape PM*, low-mass*, star, *inCl, Candidate\_low-mass*} & 17,670 (87.10\%) & 3624 (92.26\%) & 55 (83.33\%) \\
        White dwarfs & {\itshape WD*, Candidate\_WD*} & 29 (0.14\%) & 2 (0.05\%) 0 & 0 (0.00\%) \\
        Brown dwarfs & {\itshape brownD*, Candidate\_brownD*} & 45 (0.22\%) & 8 (0.20\%) & 0 (0.00\%) \\
        X-ray sources & {\itshape X} & 303 (1.49\%) & 96 (2.44\%) & 1 (1.52\%) \\
        Infrared sources & {\itshape IR,  IR<10$\mu$ m} & 1035 (5.10\%) & 92 (2.34\%) & 8 (12.12\%) \\
        Known multiple systems & {\itshape *in**, **, EB*Algol, EB*, multiple\_source, SB} & 584 (2.88\%) & 68 (1.73\%) & 1 (1.52\%) \\
        Extragalactic & {\itshape Galaxy, EmG, GinGroup, GinCl, QSO\_Candidate} & 196 (0.97\%) & 29 (0.74\%) & 1 (1.52\%)\\
        Variable stars & {\itshape V*, RotV*, Flare*, RRLyr} & 321 (1.58\%) & 7 (0.18\%) & 0 (0.00\%)\\
        Other sources & {\itshape Unknown Transient DkNeb SNR? HII Blue Symbiotic* Inexistant RGB*} & 22 (0.11\%) & 2 (0.05\%) & 0 (0.00\%) \\
        \hline
        Non contaminated sources & Blue source, Radio source, Brown dwarfs, Young stellar Objects, Infrared sources, Known multiple systems, Unknown, Potential M dwarfs, X-ray sources & 19733 (97.27\%) & 3890 (99.03\%) & 65 (98.48\%) \\
        Contaminated sources & Not an source, Symbiotic Star, ISM, White dwarfs, Extragalactic, Variable stars, Red Giant Branch Star & 553 (2.73\%) & 38 (0.97\%) & 1 (1.52\%)\\
        \hline
        \end{tabular}
        \end{center}
        \caption{Statistics on SIMBAD source classifications for the cross-match between the full catalogue of M dwarf candidates, the excess sample and the candidate M+UCDs with SIMBAD. Object classifications are only shown for those SIMBAD spectral types with non-zero cross-matches. Note some source classifications, e.g. `star', carry little information and hence contamination levels may be underestimated. \label{table:SIMBAD_type_classifications}}
    \end{table*}

    We cross-matched our full M dwarf candidate catalogue (440,694 M dwarf candidates), our excess sample (36,898 M dwarf candidates) and our M+UCD sample (1,082 M dwarf candidates) with SIMBAD. In total there were 20,286 matches with our full M dwarf candidate catalogue; 3,928 matches out of the 36,898 excess sample and 66 matches out of the 1,082 candidate M+UCDs. Of these 7,360; 1,475 and 32 respectively had spectral types from SIMBAD (See Table \ref{table:SIMBAD_spt_classifications}). We repeated this exercise with the LAMOST DR1 and DR2 catalogue spectral types. In total there were 9,262 with spectral types in our full M dwarf candidate catalogue; 1,851 with spectral types out of our 36,898 excess sample and 41 with spectral types out of our 1,082 candidate M+UCDs (see Table \ref{table:LAMOST_spt_classifications}. We also counted the source classifications given in SIMBAD and grouped them (see Table \ref{table:SIMBAD_type_classifications}).

\section{Modified reddening equation}
    \label{Appendix:modified_reddening_equation}

    \begin{table*}
       \begin{center}
        \begin{tabular}{ccccccp{2.25 cm}}
        \hline
        Colour & $E(\lambda_1-\lambda_2)$ & $A_V(CCM)$ & $A_V(F99)$ & $A_V(SFD)$ & $A_V(\bar{x})$ & No. of M dwarfs after cut \\
        \hline
        \HWa & 0.01 & 0.081 $\pm$ 0.004 & 0.093 $\pm$ 0.017 & 0.080 $\pm$ 0.001 & 0.081 $\pm$ 0.001 & ... \\
        \HWb & 0.01 & 0.069 $\pm$ 0.003 & 0.077 $\pm$ 0.005 & 0.069 $\pm$ 0.001 & 0.069 $\pm$ 0.001 & ... \\
        \JWa & 0.01 & 0.044 $\pm$ 0.003 & 0.047 $\pm$ 0.005 & 0.044 $\pm$ 0.001 & 0.044 $\pm$ 0.001 & ... \\
        \JWb & 0.01 & 0.040 $\pm$ 0.003 & 0.042 $\pm$ 0.002 & 0.041 $\pm$ 0.001 & 0.041 $\pm$ 0.001 & 13,036 \\
        \hline
        \HWa & 0.02 & 0.161 $\pm$ 0.007 & 0.187 $\pm$ 0.034 & 0.161 $\pm$ 0.003 & 0.162 $\pm$ 0.002 & ... \\
        \HWb & 0.02 & 0.137 $\pm$ 0.005 & 0.154 $\pm$ 0.011 & 0.138 $\pm$ 0.002 & 0.138 $\pm$ 0.002 & ... \\
        \JWa & 0.02 & 0.089 $\pm$ 0.005 & 0.094 $\pm$ 0.009 & 0.089 $\pm$ 0.002 & 0.089 $\pm$ 0.002 & ... \\
        \JWb & 0.02 & 0.081 $\pm$ 0.004 & 0.085 $\pm$ 0.004 & 0.081 $\pm$ 0.002 & 0.081 $\pm$ 0.001 & 45,543 \\
        \hline
        \HWa & 0.03 & 0.242 $\pm$ 0.011 & 0.280 $\pm$ 0.050 & 0.242 $\pm$ 0.004 & 0.243 $\pm$ 0.004 & ... \\
        \HWb & 0.03 & 0.206 $\pm$ 0.008 & 0.232 $\pm$ 0.016 & 0.207 $\pm$ 0.003 & 0.207 $\pm$ 0.003 & ... \\
        \JWa & 0.03 & 0.133 $\pm$ 0.008 & 0.141 $\pm$ 0.014 & 0.133 $\pm$ 0.003 & 0.133 $\pm$ 0.003 & ... \\
        \JWb & 0.03 & 0.121 $\pm$ 0.007 & 0.127 $\pm$ 0.007 & 0.122 $\pm$ 0.002 & 0.122 $\pm$ 0.002 & 69,722 \\
        \hline
        \end{tabular}
        \end{center}
        \caption{Using Equation \ref{equation:AVcut2} and $A_{\lambda}/A_V$ estimated from CCM: \protect\citet{Cardelli1989}, F99: \protect\citet{Fitzpatrick1999} and SFD: \protect\citet{Schlegel1998}. Here $R_V=3.1$ and $\bar{x}$ is the weighted average of the three estimations. Note errors from CCM are described as a lower limit only. \label{table:AV_values}}      
    \end{table*}

    We used equation \ref{equation:AVcut1} from \citet{Massa1989} and thus derived equation \ref{equation:AVcut2}, where $\frac{A_{\lambda}}{A_V}$ was calculated by taking the weighted average of cubic splines fits to $\frac{A_{\lambda}}{A_V}(\lambda^{-1})$ from \citet{Cardelli1989}, \citet{Fitzpatrick1999} and \citet{Schlegel1998} for an $R_V$ of 3.1. Note we tested $R_V$ values of 2.1 , 3.1 and 4.1 (See Table \ref{table:AV_values}). At these tiny values of extinction an $R_V$ value of 3.1 is satisfactory.

    \begin{equation}
        \label{equation:AVcut1}
        A_{\lambda1} - A_{\lambda2} = E(\lambda_1-\lambda_2) \\
    \end{equation}
        
    \begin{equation}
        \label{equation:AVcut2}    
        A_V \leqslant E(\lambda_1-\lambda_2)\left[\frac{A_{\lambda1}}{A_V}-\frac{A_{\lambda2}}{A_V}\right]^{-1}
    \end{equation}

\section{Photometric simulation}
    \label{Appendix:simulation}

        A polynomial was fit to the data points using a Bayesian approach (using {\sc emcee}\footnote{pure-{\sc python} implementation of \citet{Goodman2010} affine invariant Markov Chain Monte Carlo ensemble sampler} and the fitting routine used by \citealt{Foreman-Mackey2013} and \citealt{Hogg2010}).

        The probabilistic fitting routine allowed the polynomial parameters ($a_i = a_1, a_2, ..., a_n$) to vary as well as allowing the variance to vary\footnote{See \url{http://dan.iel.fm/emcee/current/user/line/} for a full example}, represented below by $f$. The probability distribution is assumed Gaussian and is shown in equation \ref{equation:mcmc}. 

        \begin{equation}
            \label{equation:mcmc}
            \ln p(y | x, \sigma, model, f) = \\ -\frac{1}{2} \sum_{i=1}^{n} \frac{(y_n-model_n)^2}{s_n^2} + \ln(2\pi s_n^2)
        \end{equation}
        \\
        where $s_n^2 = \sigma_n^2 + f^2(model_n)^2$ and $model_n = \sum_{i=0}^{m} a_i x^i_n$.

\begin{figure*}
            \begin{center}
            \includegraphics[width=0.8\textwidth]{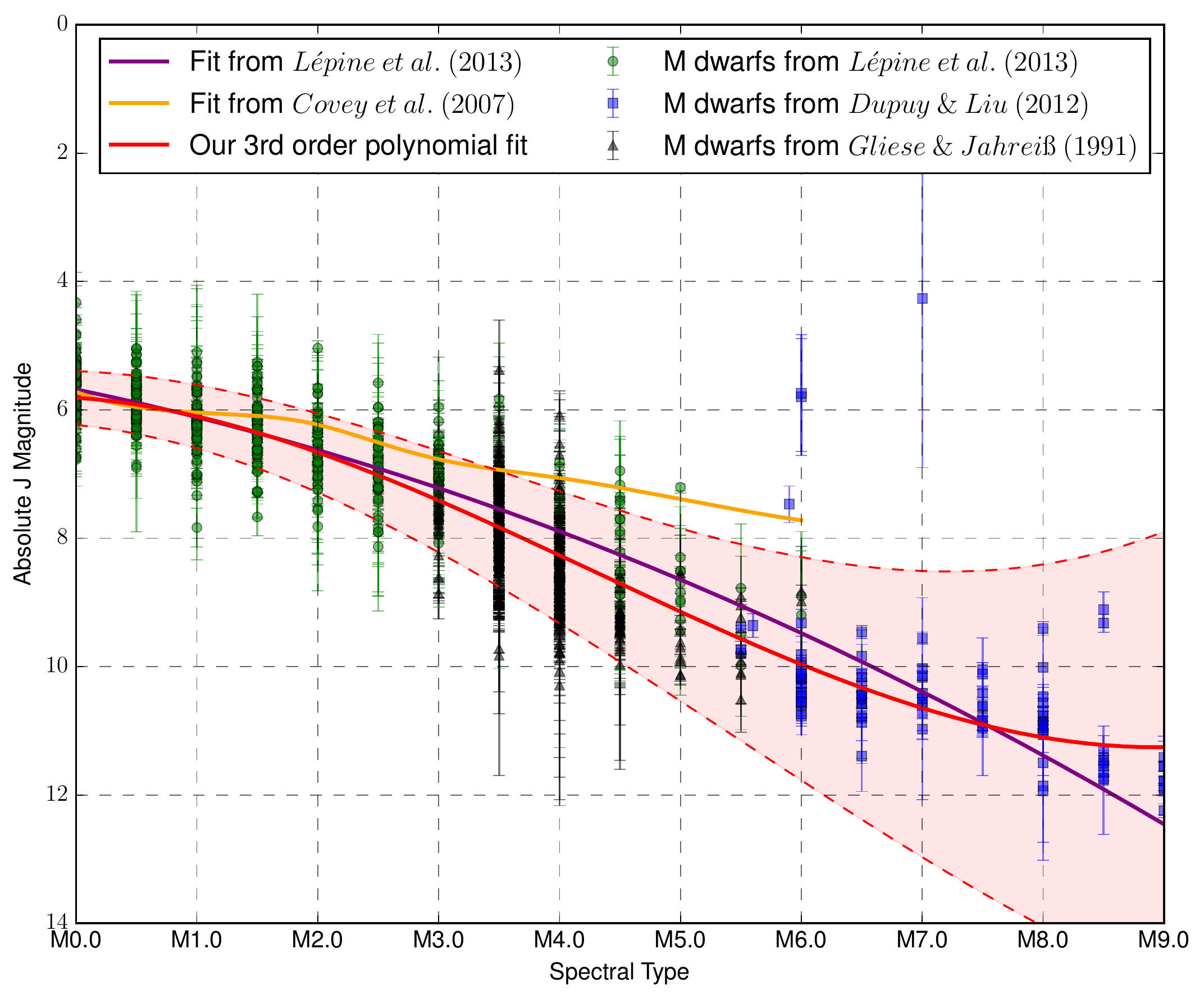}
            \caption{Absolute magnitude against spectral type for \protect\citet{Lepine2013, Gliese1991} and \protect\citet{Dupuy2012} M dwarfs. Shown in purple is the fit \protect\citet{Lepine2013} proposed, in yellow is an interpolated fit of data from \protect\citet{Covey2007} and in red is our spectral type fit (see equation \protect\ref{equation:MJ_spt}), shaded regions show outer most bounds of the uncertainties and the added uncertainty due to the contribution of an unresolved equal binary ($\pm$0.375 orders of magnitude$^{\protect\ref{footnote:magnitudeofanequalbinary}}$). \label{figure:Mdwarf_fit}}
            \end{center}
        \end{figure*}

        The best polynomial fit found to simulate absolute J band magnitude, \MJ, from spectral subtype, $spt$, for spectral subtype in the range M1 $\geq spt \geq$ M8 was a cubic fit (equation \ref{equation:MJ_spt}). 

        \begin{equation}
            \label{equation:MJ_spt}
            \MJ = \begin{bmatrix}-(0.014^{+0.002}_{-0.002})spt^3 + (0.17^{+0.02}_{-0.02})spt^2 + \\\\ (0.13^{+0.05}_{-0.06})spt + (5.81^{+0.05}_{-0.04})\pm0.375 \end{bmatrix}
        \end{equation}
        \\
        where the $\pm0.375$ is added to simulate the maximum deviation due to binaries in our sample\footnote{The maximum brightness of an unresolved binary for two stars of equal brightness giving a factor of two in flux (in magnitudes equivalent to $-2.5\log_{10}(2) \approx -0.75 \rightarrow \pm0.375$ uncertainty). \label{footnote:magnitudeofanequalbinary}} (see Figure \ref{figure:Mdwarf_fit}). This enabled the primary and companion sensitivity to be modelled for all combinations of colour (Figures \ref{figure:expectedexcess} and \ref{figure:expected_colour_sense}).

        Using spectra from the SpeX Prism Spectral Libraries\footnote{SpeX Prism Spectral Libraries, maintained by Adam Burgasser at \url{http://pono.ucsd.edu/~adam/browndwarfs/spexprism}. \label{footnote:Spex}} we combined M dwarf and UCD near-infrared spectra to simulate M dwarf + UCD unresolved binary systems. From the spectra of the M dwarfs and of the M dwarf + UCD unresolved binary systems the contribution due to the addition of a UCD was calculated. This figure compliments the simulated photometric excesses in Figure \ref{figure:colour_sense}, note the excesses in Figure \ref{figure:colour_sense} are the mean colour excess across M3 to M6 and L0 to L4, and thus appear diluted when compared to the peak excess (around L2). The peak excess around L2 is also seen in Figure \ref{figure:expectedexcess}b thus validating our photometric simulations spectroscopically.

        \begin{table*}
            \begin{minipage}{\textwidth}
            \begin{center}
            \begin{tabular}{cccccc}
            \hline
            y & x & $c_0$ & $c_1$ & $c_2$ & $c_3$ \\
            \hline 
            \MJ& $spt$& $(+5.81^{+0.05}_{-0.04})\pm0.375$ & $(+1.28^{+0.55}_{-0.55})\tx^{-1}$ & $(+1.76^{+0.20}_{-0.17})\tx^{-1}$ & $(-1.37^{+0.16}_{-0.15})\tx^{-2}$ \\
            \JH  & \VJ   & $(+9.20^{+0.37}_{-0.36})\tx^{-1}$         & $(-1.38^{+2.55}_{-2.93})\tx^{-1}$ & $(+1.33^{+0.23}_{-0.23})\tx^{-2}$ & ... \\
            \JK  & \VJ   & $(+9.18^{+0.41}_{-0.36})\tx^{-1}$         & $(-0.59^{+0.99}_{-1.38})\tx^{-1}$ & $(+0.96^{+0.25}_{-0.22})\tx^{-2}$ & ... \\
            \JWa & \VJ   & $(+8.05^{+0.91}_{-0.74})\tx^{-1}$         & $(+4.05^{+4.26}_{-4.01})\tx^{-2}$ & $(+0.30^{+0.55}_{-0.12})\tx^{-2}$ & ... \\
            \JWb & \VJ   & $(+1.74^{+1.27}_{-1.01})\tx^{-1}$         & $(+3.58^{+0.56}_{-0.59})\tx^{-1}$ & $(+2.76^{+4.77}_{-6.18})\tx^{-2}$ & ... \\
            \hline
            \end{tabular}
            \end{center}
            \caption{M dwarf Synthetic Photometry Generation: Colour-colour fits ($y = \sum_{i=0}^{N} c_ix^i$), where x=0 corresponds to M0 and x = 6 corresponds to M6, were produced for M dwarfs M0 - M6 using our probabilistic fitting routine (equation \ref{equation:mcmc}) using the data from \protect\citet{Lepine2013, Gliese1991} and \protect\citet{Dupuy2012}. \label{table:Sim_MD}}
            \end{minipage}
        \end{table*}

        \begin{table*}
            \begin{minipage}{\textwidth}
            \begin{center}
            \begin{tabular}{cccccc}
            \hline
            y   & $c_0$ & $c_1$ & $c_2$ & $c_3$ \\
            \hline
            \UG & $(+6.29^{+7.23}_{-2.74})\tx^{+0}$ & $(-1.02^{+1.59}_{-3.32})\tx^{+0}$ & $(+5.24^{+8.37}_{-1.93})\tx^{-2}$ & $(-0.09^{+0.15}_{-0.33})\tx^{-2}$ \\
            \GR & $(+9.87^{+5.26}_{-8.81})\tx^{-1}$ & $(+2.61^{+5.04}_{-0.69})\tx^{-1}$ & $(-0.98^{+1.71}_{-4.16})\tx^{-2}$ & ... \\
            \RI & $(+2.74^{+0.32}_{-0.27})\tx^{+0}$ & $(+8.10^{+4.40}_{-2.54})\tx^{-2}$ & $(-0.23^{+0.31}_{-0.59})\tx^{-2}$ & ... \\
            \IZ & $(+1.70^{+0.65}_{-0.55})\tx^{+0}$ & $(-0.90^{+0.16}_{-0.31})\tx^{-1}$ & $(+1.31^{+1.10}_{-0.97})\tx^{-2}$ & $(-0.03^{+0.03}_{-0.08})\tx^{-2}$  \\
            \ZJ & $(+2.08^{+9.40}_{-0.86})\tx^{-1}$ & $(+4.05^{+1.48}_{-1.06})\tx^{+0}$ & $(-2.15^{+3.24}_{-5.02})\tx^{-2}$ & $(+0.04^{+0.02}_{-0.02})\tx^{-2}$  \\
            \hline
            \end{tabular}

            \end{center}
            \caption{UCD Synthetic Photometry Generation: $u$, $g$, $r$, $i$ and $z$ were generated using the probabilistic fitting routine on data from \protect\citet{Hawley2002, Chiu2006} and \protect\citet{Dupuy2012} cross-matched with 2MASS and SDSS and J from \citet{Dupuy2012}, where $y = \sum_{i=0}^{N} c_{i}x^{i}$, and $x$ is spectral type. \label{table:Sim_BD}} 
            \end{minipage}
        \end{table*}

        \begin{figure*}
            \begin{center}
            \includegraphics[width=0.8\textwidth]{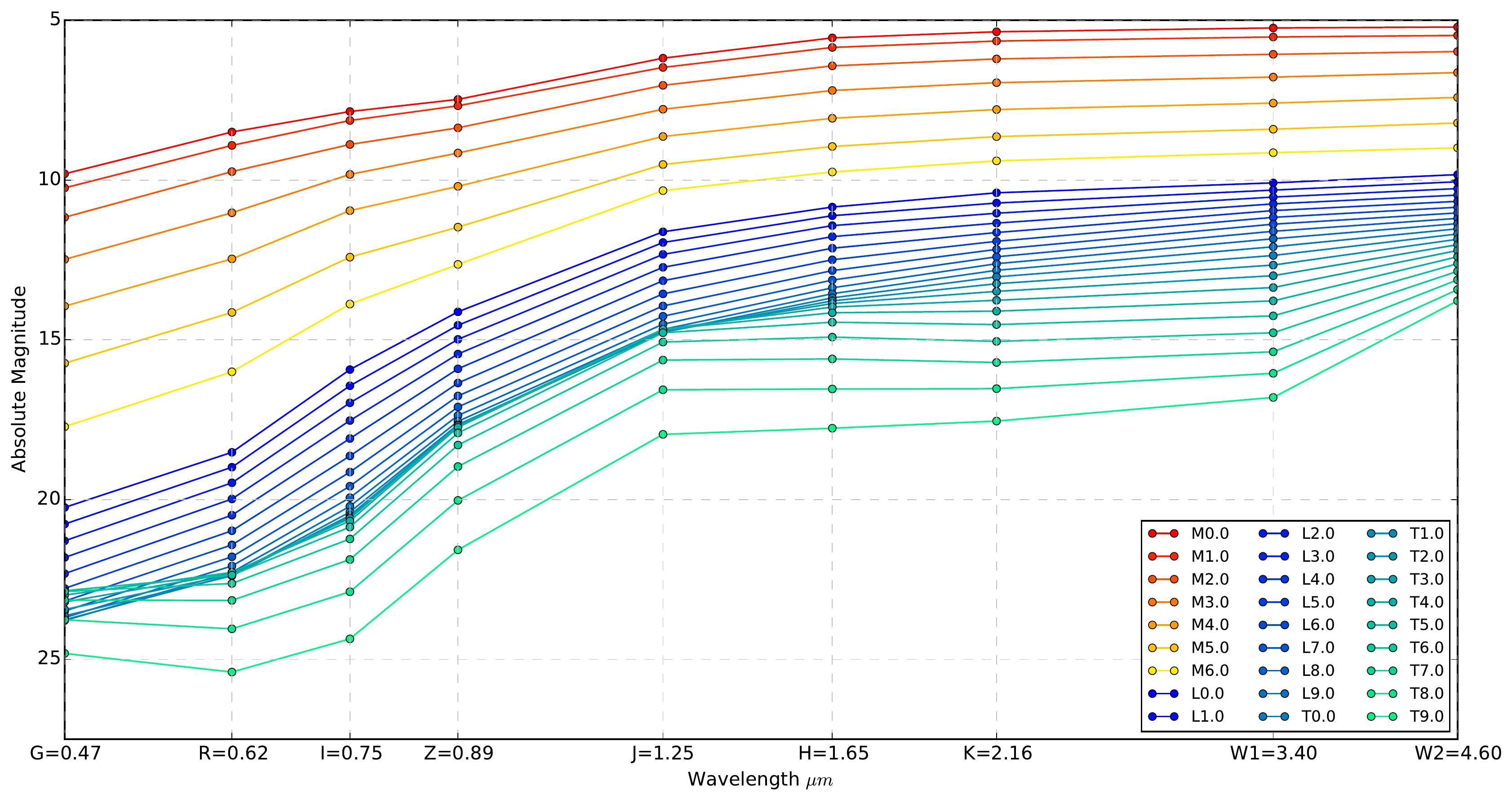}
            \caption{ Absolute magnitude against wavelength, simulated M dwarf and UCD photometry as calculated by the fits in Table \protect\ref{table:Sim_MD} and Table \protect\ref{table:Sim_BD}. From our fit we include shaded regions bounded by dashed red line to show the uncertainties of our fit. \label{figure:SED}}
            \end{center}
        \end{figure*}
       
        \begin{figure*}
            \begin{center}
            \begin{minipage}{.48\textwidth}
                \begin{center}
                \includegraphics[width = 8.0 cm]{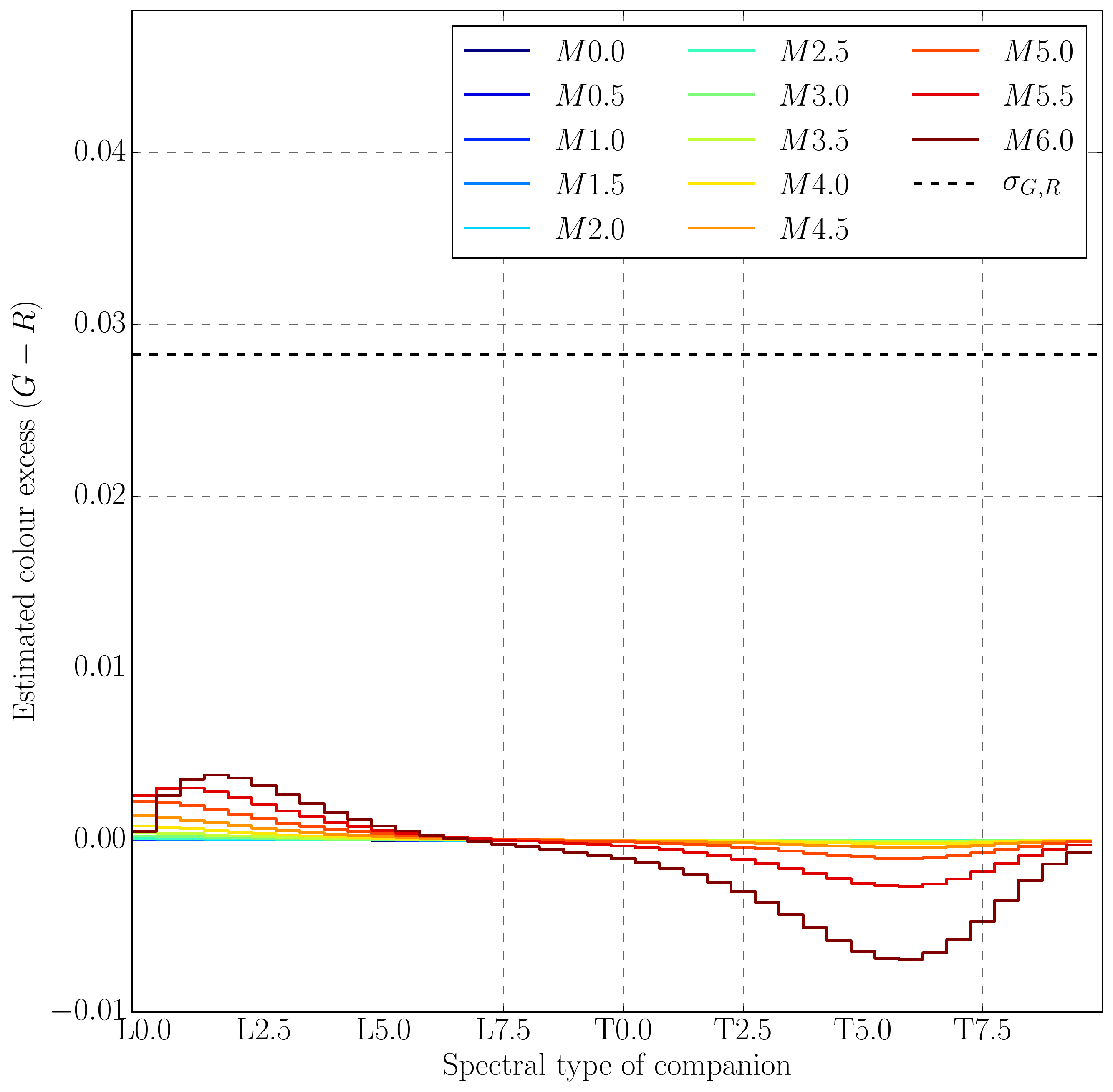}
                \vspace{0.1cm}
                \\(a)
                \vspace{0.1cm}
            \end{center}
            \end{minipage}
            \begin{minipage}{.48\textwidth}
                \begin{center}
                \includegraphics[width = 8.0 cm]{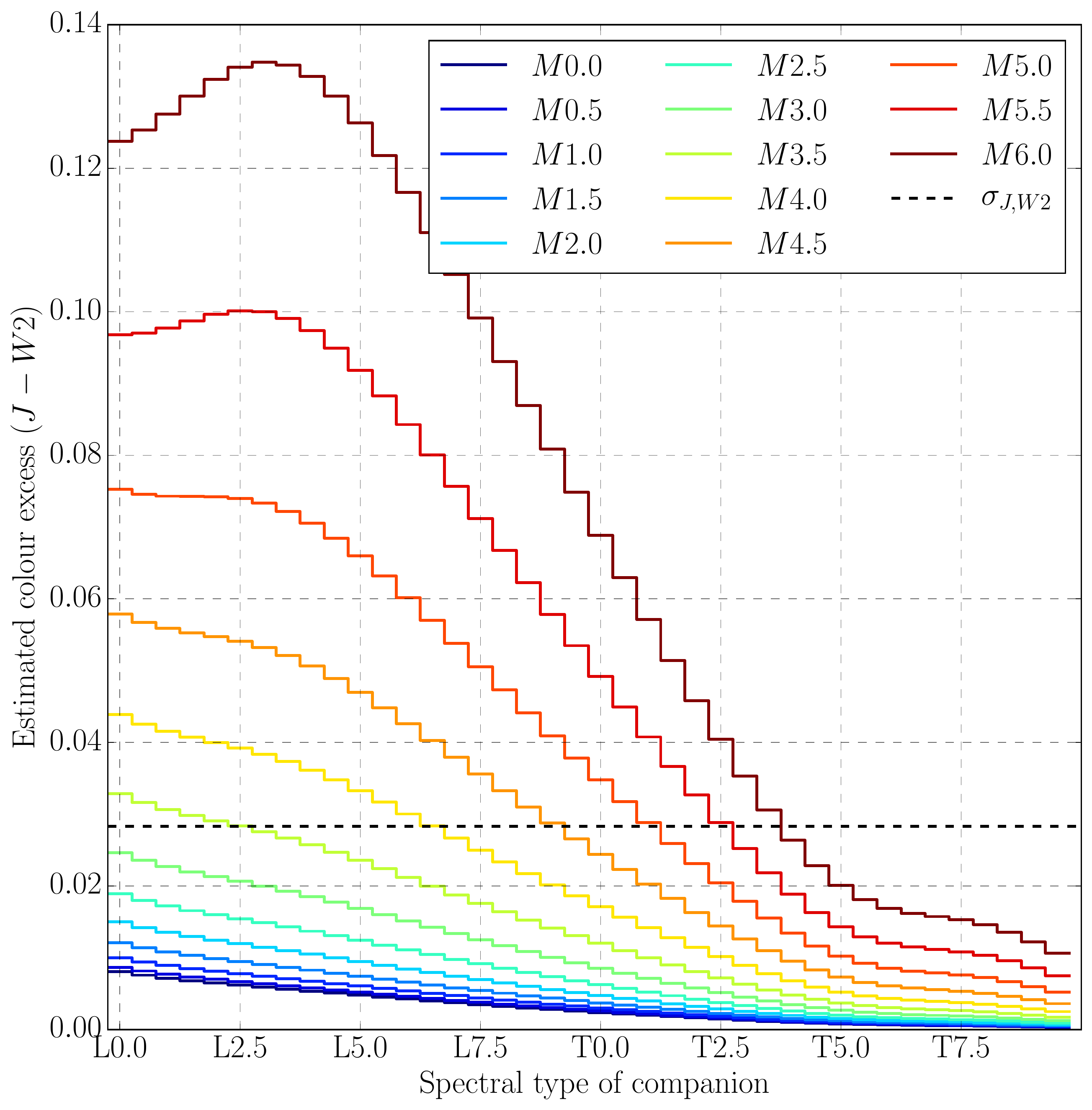}
                \vspace{0.1cm}
                \\(b)
                \vspace{0.1cm}
                \end{center}
            \end{minipage}
            \begin{minipage}{\textwidth}
                \caption{ Expected colour excess for simulated M0 + UCD to M6 + UCD for (a) \GR : an identified PS colour and (b) \JWb: an identified CS colour. Note the difference in scales of the colour excess (y) axis. The dashed line shows the average colour uncertainties for these colours in our catalogue. \label{figure:expectedexcess}}
            \end{minipage}\qquad
            \end{center}
        \end{figure*}

        \begin{figure*}
            \begin{center}
            \begin{minipage}{.48\textwidth}
                \begin{center}
                \includegraphics[width = 8.0 cm]{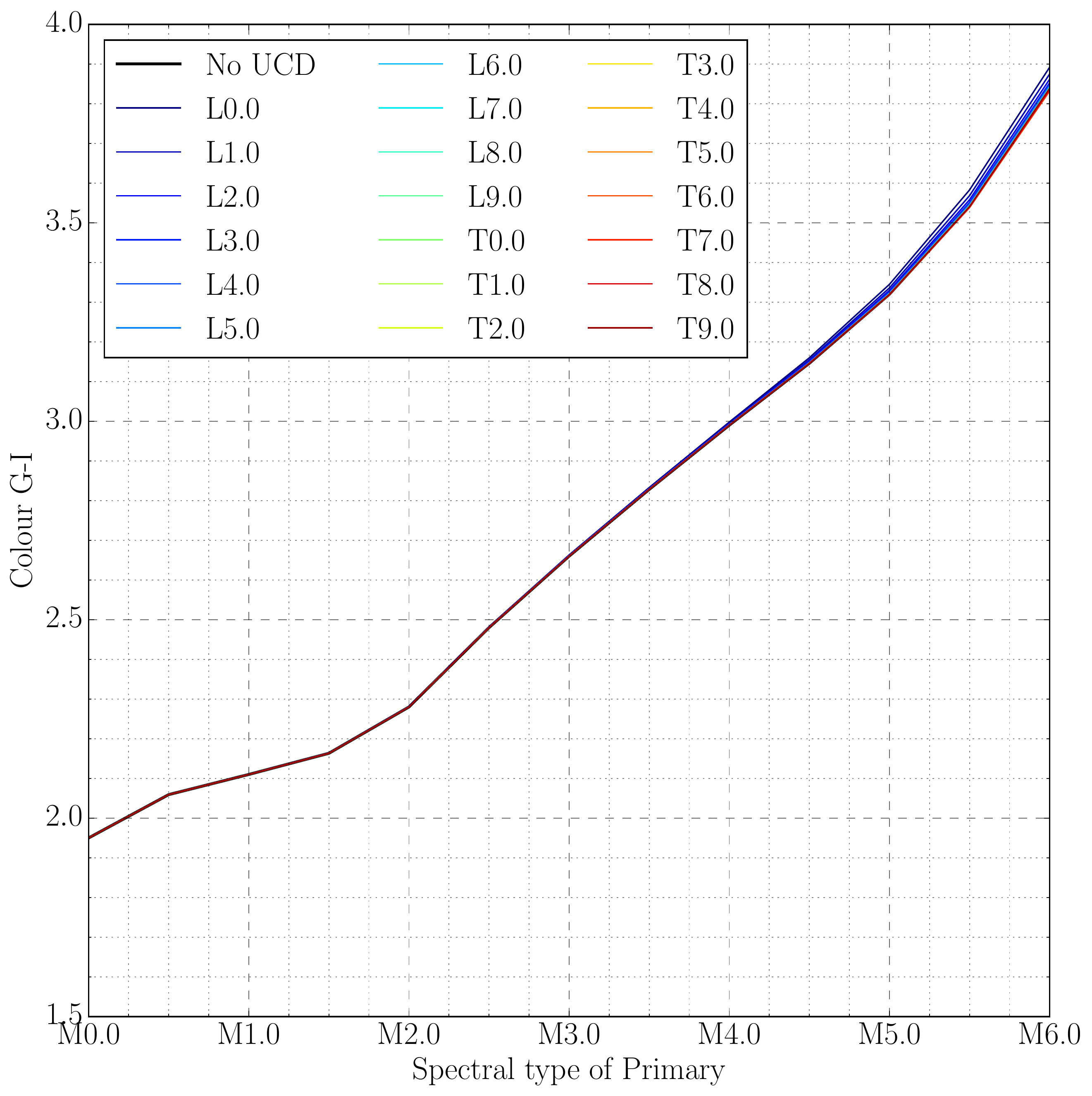}
                \vspace{0.1cm}
                \\(a)
                \vspace{0.1cm}
                \end{center}
            \end{minipage}
            \begin{minipage}{.48\textwidth}
                \begin{center}
                \includegraphics[width = 8.0 cm]{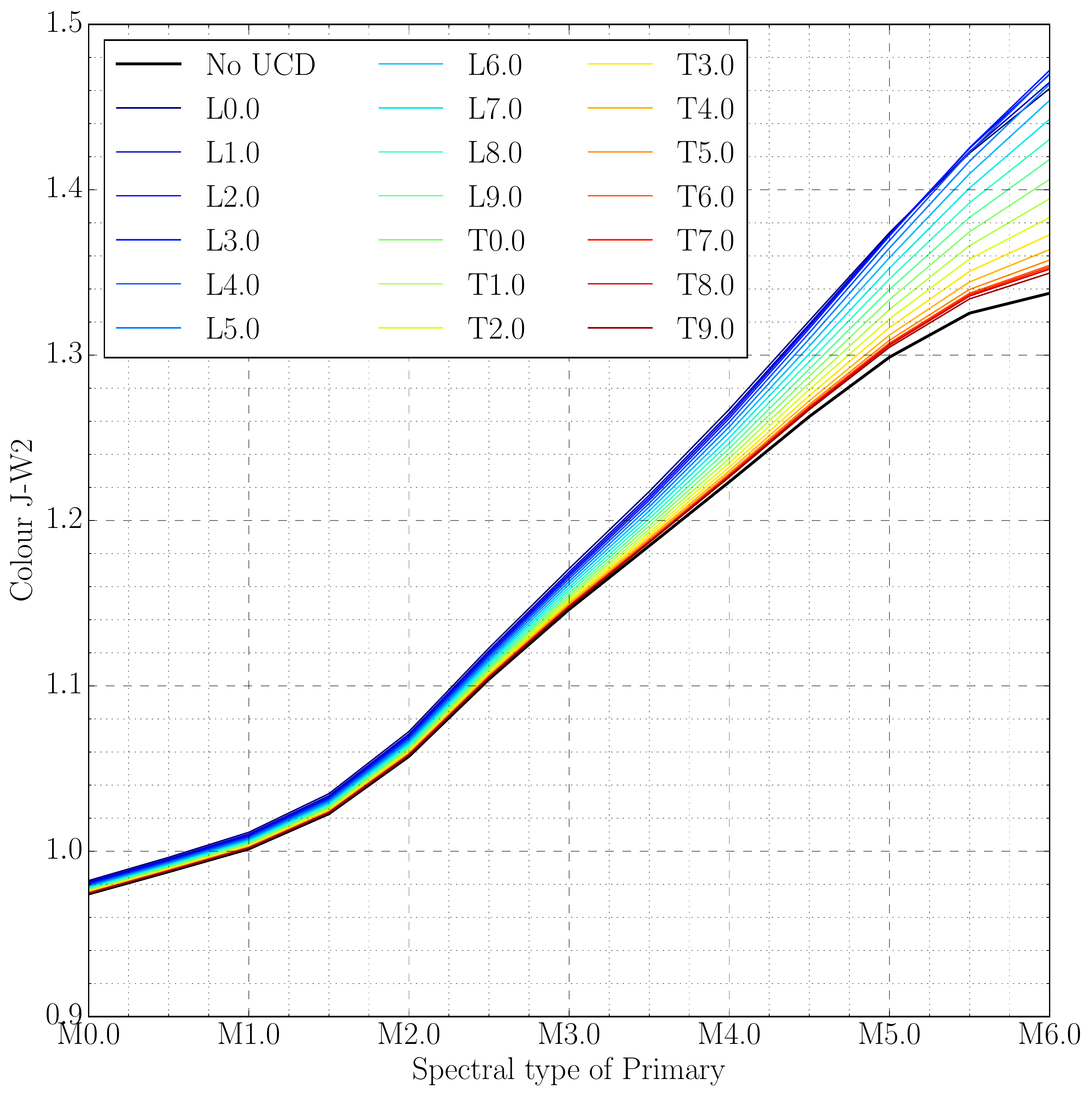}
                \vspace{0.1cm}
                \\(b)
                \vspace{0.1cm}
                \end{center}
            \end{minipage}
            \begin{minipage}{\textwidth}
            \caption{ Expected primary colour variation in (a) \GI and (b) \JWb as a function of primary spectral type for simulated M dwarf+UCD unresolved binary systems. A large change in colour with primary spectral type without a large change across companion spectral type identifies a good colour for identifying M dwarfs. Note the difference in scales of the colour (y) axis. \label{figure:expected_colour_sense}}
            \end{minipage}
            \end{center}
        \end{figure*}

        \begin{figure*}
            \begin{center}
                \includegraphics[width=0.75\textwidth]{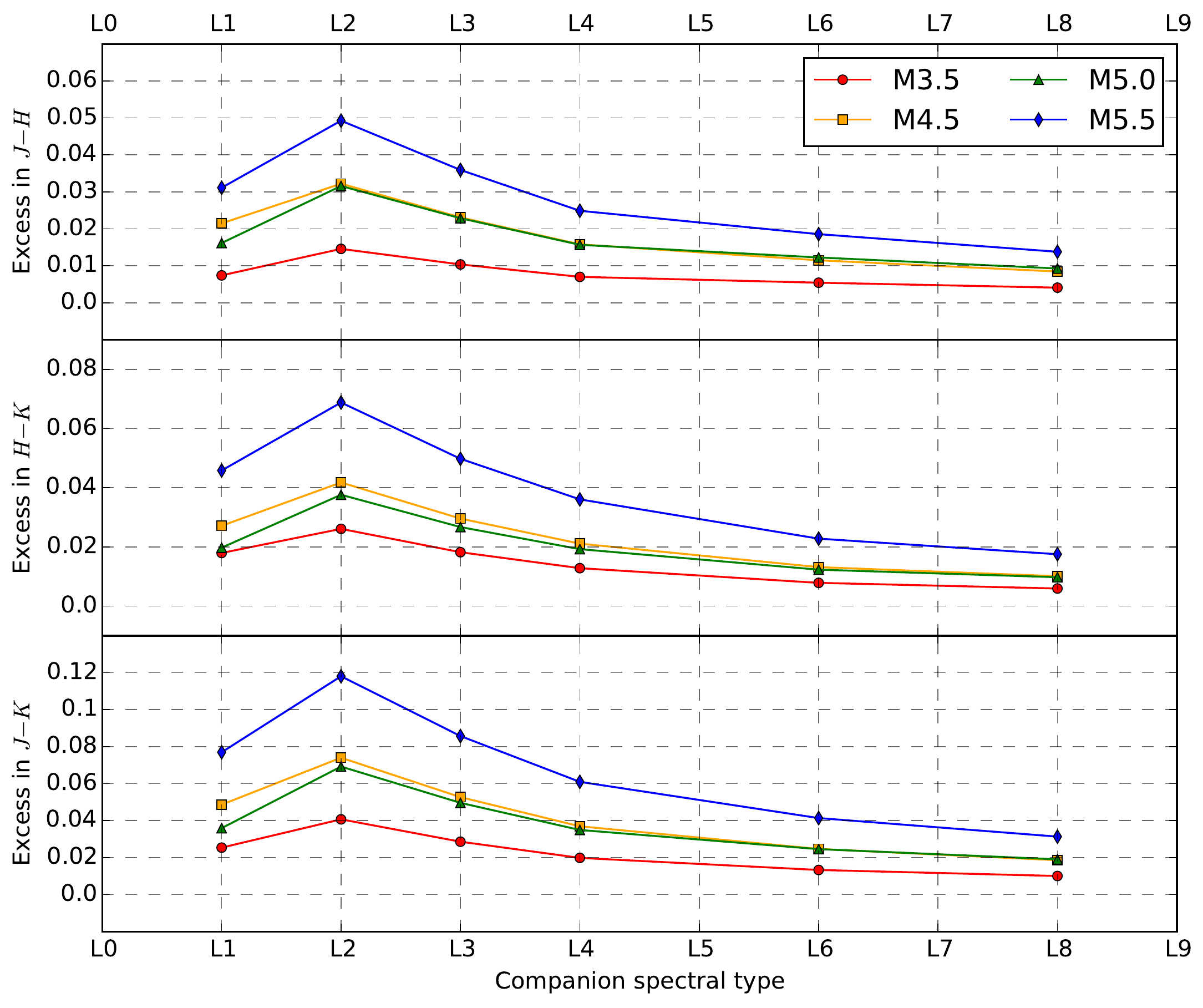}
            \caption{Using spectra from SpeX$^{\protect\ref{footnote:Spex}}$ we combined M dwarf and UCD near-infrared spectra to simulate M dwarf+UCD unresolved binary systems. We calculated the \JH, \HK and \JK colours for each and compared them to the colours of the isolated M dwarfs. The excess seen is complimentary to the photometric simulations (Figure \protect\ref{figure:colour_sense} where we average over M3 to M6 and L0 to L4). \label{figure:colour_from_spec}}
            \end{center}
        \end{figure*}

\bsp	% typesetting comment

\label{lastpage}
\end{document}